\documentclass[aps,nofootinbib,floatfix,prd,letterpaper,twocolumn,superscriptaddress]{revtex4}
\usepackage{graphicx,amsmath,amssymb,color}
\def\be{\begin{equation}}
\def\ee{\end{equation}}
\def\bea{\begin{eqnarray}}
\def\eea{\end{eqnarray}}
\newcommand{\vs}{\nonumber\\}
\def\ba#1\ea{\begin{align}#1\end{align}}

\newcommand{\comment}[1]{}

\newcommand{\g}{\gamma}
\newcommand{\s}{\sigma}
\newcommand{\refeq}[1]{Eq.~(\ref{eq:#1})}          
\newcommand{\refeqs}[2]{Eqs.~(\ref{eq:#1})--(\ref{eq:#2})}          
          
\newcommand{\reffig}[1]{Fig.~\ref{fig:#1}}          
\newcommand{\reffigs}[2]{Figs.~\ref{fig:#1}--\ref{fig:#2}}          
          
\newcommand{\refsec}[1]{Sec.~\ref{sec:#1}}          
\newcommand{\refapp}[1]{App.~\ref{app:#1}}
\newcommand{\reftab}[1]{Tab.~\ref{tab:#1}}

\renewcommand{\v}[1]{\mathbf{#1}}
\newcommand{\vx}{\v{x}}
\renewcommand{\vr}{\v{r}}
\newcommand{\vk}{\v{k}}
\newcommand{\vq}{\v{q}}

\newcommand{\<}{\langle}
\renewcommand{\>}{\rangle}

\renewcommand{\k}{\kappa}
\renewcommand{\d}{\delta}
\newcommand{\D}{\Delta}

\newcommand{\nhat}{\hat{n}}
\newcommand{\vnhat}{\v{\hat{n}}}

\newcommand{\vn}{\v{\nabla}}
\newcommand{\Om}{\Omega_m}

\newcommand{\Omn}{\Omega_{m0}}

\newcommand{\zt}{\tilde{z}}

\newcommand{\chib}{\bar{\chi}}
\newcommand{\chit}{\tilde{\chi}}
\newcommand{\aeq}{a_{\rm eq}}
\def\Q{\mathcal{Q}}

\def\M{\mathcal{M}}

\def\H{\mathcal{H}}

\def\P{\mathcal{P}}

\def\O{\mathcal{O}}

\def\A{\mathcal{A}}
\def\B{\mathcal{B}}
\def\C{\mathcal{C}}
\def\T{\mathcal{T}}

\def\iMpch{\,h\, {\rm Mpc}^{-1}}
\newcommand{\Tr}{{\rm Tr}\,}

\newcommand{\fncb}{$\overline{\mathrm{FNC}}$}
\newcommand{\fnc}{$\mathrm{FNC}$}

\def\uu{\upsilon}

\newcommand{\cH}{\mathcal{H}}
\newcommand{\ebf}{\bar \tau_{F}}

\def\nhat{\hat{n}}
\def\vnhat{\hat{\v{n}}}

\begin{document}

\title{Large-Scale Structure and Gravitational Waves III: Tidal Effects}

\author{Fabian Schmidt}
\affiliation{Max-Planck-Insitute for Astrophysics, D-85748 Garching, Germany}
\affiliation{Department of Astrophysical Sciences, Princeton University,
Princeton, NJ~08544, USA}

\author{Enrico Pajer}
\affiliation{Department of Physics, Princeton University, Princeton, NJ~08544, USA}

\author{Matias Zaldarriaga}
\affiliation{Institute for Advanced Study, Princeton, NJ~08540, USA}

\begin{abstract}
The leading locally observable effect of a long-wavelength metric
perturbation corresponds to a tidal field.  We derive the tidal
field induced by scalar, vector, and tensor perturbations, and use
second order perturbation theory to calculate the
effect on the locally measured small-scale density fluctuations.  
For sub-horizon scalar perturbations, we recover the standard 
perturbation theory result ($F_2$ kernel).  
For tensor modes of wavenumber $k_L$, we find that effects persist
for $k_L\tau \gg 1$, i.e.~even long after the gravitational
wave has entered the horizon and redshifted away, i.e.~it is a ``fossil'' effect. 
We then use these results, combined with the ``ruler perturbations'' of
\cite{stdruler}, to predict the observed distortion of the small-scale
matter correlation function induced by a long-wavelength tensor mode.  We also
estimate the observed signal in the B mode of the cosmic shear from
a gravitational wave background, including both tidal (intrinsic alignment) and projection (lensing) effects.
The non-vanishing tidal effect in the $k_L\tau \gg 1$ limit significantly
increases the intrinsic alignment contribution to shear B modes, 
especially at low redshifts $z \lesssim 2$.
\end{abstract}

\date{\today}

\maketitle

%%%%%%%%%%%%%%%%%%%%%%%%%%%%%%%%%%%%%%%%%%%%%%%%%%%%%%%%%%%%%%%%%%%%%%%%%%%
%%%%%%%%%%%%%%%%%%%%%%%%%%%%%%%%%%%%%%%%%%%%%%%%%%%%%%%%%%%%%%%%%%%%%%%%%%%
\section{Introduction}
\label{sec:intro}

Cosmological perturbation theory is a robust pillar on which our interpretation of cosmological observations rests. Although the linear results are by now part of textbook material, second and higher order effects have not yet been comprehensively computed. Given the ever increasing amount and precision of observations, many of these effects have already been or soon will be detected, providing strong motivation for a growing body of work (see \cite{2013JCAP...11..015C,PietroniPeloso,Creminelli:2013poa,Kehagias1,Valageas,Creminelli:2013mca,Bartolo:2010rw,Giddings:2011zd,Kehagias2,Kehagias3} for recent developments). 

While independent in linear perturbation theory, scalar, vector, and tensor 
modes are coupled at second order.  This leads to interesting effects which
have only recently been begun to be explored.  While vector modes decay at linear order also on super horizon scales, tensors are conserved and might therefore have survived since the very early universe, thereby providing us with a unique opportunity to peek almost directly into those early stages of cosmological evolution.  Specifically, a measurement of a scale-invariant
background of gravitational waves would provide strong support for the
inflationary paradigm, and tell us about the energy scale of inflation.

In this paper, we show that the leading locally observable effect of a 
long-wavelength perturbation $k_L$ (be it scalar, vector, or tensor) on small-scale
density perturbations with $k_{S}\gg k_{L}$ is given by an effective tidal field, and we derive the resulting contribution to the density field at second order.  
Our formalism thus captures any purely gravitational coupling at leading order
in $k_L/k_S$.   
The most well-known case is a long-wavelength density (scalar) perturbation, whose
effect on small-scale fluctuations is given by standard second-order 
perturbation theory (specifically the $F_2$ kernel).  As a check of
our formalism and computations, we re-derive this standard result. Our results are new for tidal fields of vector origin, which, to the best of our knowledge, have not been previously considered in the literature. For the case in which the tidal field is generated by gravitational waves (tensor modes), first estimates of the tidal effects were given in \cite{MasuiPen,GWshear}. More recently, a detailed computation has been presented in \cite{dai/etal:13} assuming matter domination. Comparison with these previous works is in order. In \cite{MasuiPen}, the anisotropy in the short-scale power spectrum was estimated to be of order $h^{(0)}_{ij}k_{S}^{i}k_{S}^{j}/k_{S}^{2}$, where $h^{(0)}_{ij}$ is the spatial metric perturbation at early times.  Although this agrees with our final result and that of \cite{dai/etal:13} up to factors of order one, it does not capture the  time dependence of the effect which sheds light on the physical origin of the effect as we will see momentarily.  In \cite{GWshear} it was roughly estimated that the ``intrinsic'' shape correlations of galaxies induced by tensor modes are proportional to the instantaneous amplitude of the tensor mode tidal field, while we will argue that it should be more accurately given by a certain time integral over the tidal field.  Let us defer galaxy alignments for the moment and consider the anisotropy of small-scale density statistics. 
We reproduce the results of \cite{dai/etal:13} in the matter dominated regime ($k_L \lesssim 0.01\iMpch$).  However, we argue that a signal of comparable size will come from smaller scale tensor modes that entered the horizon during radiation domination.  Our treatment of radiation domination neglects perturbations in radiation, which we will study in a separate publication. This additional effect will however not change the above conclusion.  Let us stress that by dividing the effects into separately observable pieces as we will describe now, we believe our approach makes the physics of the tensor mode effects intuitive and transparent.  

Before presenting our results, we briefly summarize our methodology.  The purely gravitational effect of long wavelength metric perturbations  on short scales can be captured in a convenient and physically transparent way adopting a series of different coordinates \cite{conformalfermi}. Given some set of long-wavelength primordial metric perturbations $h(k_{L})$, one can define \textit{conformal Fermi Normal Coordinates} (\fncb) at all times along any chosen timelike geodesic. In these local coordinates, the metric is FLRW along the central geodesic with all physical effects due to $h(k_{L})$ being encoded in corrections to the metric at order $(\nabla\nabla h)\,x^2$. Results obtained in these coordinates have a clear physical interpretation as corresponding to what a local freely falling observer moving along the central geodesic would measure. 

\begin{table*}
\centering
\begin{tabular}{|c|l|l|}  \hline
Symbol    & Relation &	Meaning 	\\ \hline\hline
$\cH$ & = $a' / a = a H$ & comoving Hubble scale \\ \hline
$\delta$ & $ =\rho_{m}(\vx)/\bar \rho_{m}-1 $ & matter density perturbation  \\ \hline
$\Phi_{s}$ & \refeq{FNCmetric} & Newtonian gauge potential in absence of tidal field\\ \hline
$t_{ij}, T$ &$t_{ij}(\v{0};\tau) = T(k_L,\tau)\, t_{ij}^{(0)}(\v{0})$& tidal field $t_{ij}$ and its transfer function $T$ \\ \hline
$\alpha, \beta, \gamma$ &\refeq{d2tLPT2} & functions of $k$ and $\tau$  \\ \hline
$V,F,D_{\sigma_{1}},D_{\sigma_{2}}$ &\refeq{Fdef}, \refeq{Vdef}, \refeq{Dsigma1}, \refeq{Dsigma2} & functions of $k$ and $\tau$  \\ \hline
$\v{s}$ & $\vx(\v{q},\tau) = \v{q} + \v{s}(\v{q},\tau)$ & Lagrangian displacement vector \\ \hline
$\v{q}$ & $\vx(\v{q},\tau) = \v{q} + \v{s}(\v{q},\tau)$ & Lagrangian coordinate  \\ \hline
$\v{x}$ & $\vx(\v{q},\tau) = \v{q} + \v{s}(\v{q},\tau)$ & Eulerian coordinate \\ \hline
$u^i$ & \refeq{udef} & peculiar velocity \\ \hline
$\theta$ & $= \partial_{i} u^{i}$ & peculiar velocity divergence \\ \hline
subscripts $1,2,\dots$ & $\delta= \delta_{1}+\delta_{2}+\dots$ & (matter) perturbations at linear, second, ... order \\ \hline
subscripts $t$ or $s$ & $\delta_{1}=\delta_{1,s}+\delta_{1,t}$ & 
contributions at 0'th and first order in the tidal field
\\ \hline
subscripts $L$ or $S$ & $k_{L}\ll k_{S}$ & Long $L$ and Short $S$ wavelengths\\ \hline
superscript $(0)$ & $h_{ij}^{(0)}=h_{ij}(\tau_{0})$ & quantity evaluated at ``initial'' time $\tau_{0}$\\ \hline
\end{tabular}
\caption{Symbols used in the paper. \label{tab:not}}
\end{table*}

In the end, one is typically interested in how the perturbations $h(k_L)$ affect measurements by an observer far away, e.g.~on Earth.  This requires computing \textit{projection effects}, which we define as the mapping from local \fncb~coordinates to the coordinates chosen by the observer.  We will perform this mapping using the results of \cite{stdruler}. An advantage of this methodology is that all results at all points of the computation represent physical observables. This is to be contrasted with using global coordinates, in which case unphysical contributions from different parts of the computation need to cancel each other before the final observable result is obtained.

Our main result for tensor (and vector) perturbations can be summarized as 
follows.  Given a primordial spatial metric perturbation\footnote{For vector perturbations, we here assume instantaneous generation at some initial time.} $h_{ij}^{(0)}$ at
position $\vx$ (traceless by assumption) with comoving wavenumber $k_L$, the density 
field $\d_{2,t}$ at conformal time $\tau$ is given in terms of the linear density 
field $\d_{1,s}$ at the same time by
\ba
& \d_{2,t}(\vx,\tau) \label{eq:d2tintro}\\
& \quad = h_{ij}^{(0)}(\vx) \left[ \alpha(k_L,\tau) \frac{\partial^i\partial^j}{\nabla^2} + \beta(k_L,\tau) x^i \partial^j \right] \d_{1,s}(\vx,\tau)\,.
\nonumber
\ea
The notation $\d_{2,t}$ indicates that this is the correction to the linear
density field induced at second order by vector and tensor modes (see 
\reftab{not} for a summary of our notation).  
Here the coefficients $\alpha,\,\beta$ are functions of the wavenumber $k_L$
and conformal time $\tau$.  In particular, for $k_L\tau \ll 1$, i.e. when
the long-wavelength modes are superhorizon, $\alpha$ and $\beta$ go to zero
as demanded by the equivalence principle.  More interesting is the opposite limit
$k_L\tau \gg 1$, when the tensor or vector mode has long entered the horizon
and decayed.  
We will see that, despite what one might have expected, the coefficients
$\alpha$ and $\beta$ do not vanish in this limit but rather asymptote
to constant values $\alpha_\infty,\,\beta_\infty$. Thus, the
small-scale density field preserves the knowledge of the primordial vector
or tensor modes which have long decayed away. Such a feature was called
\emph{fossil effect} in \cite{MasuiPen}, and we will adopt this name for
the $k_L\tau \gg 1$ limit of the tidal effects.  Notice that the fossil effect is generated at the time of horizon re-entry of the tensor or vector mode. For modes of observational interest, this happens at $z<10^{5}$, much after inflation and reheating. In this sense, although the size of the effect is proportional to the primordial tensor modes, the physical coupling between tensors and scalars that we discuss in this work is generated in the late universe.  

The two terms in \refeq{d2tintro} correspond to different physical effects.  
The first term ($\propto\alpha$) indicates the effect of the tidal field on the evolution
of small-scale fluctuations. The second term ($\propto\beta$) on the other hand
encodes the effect of the displacement of matter by the long-wavelength tidal field.  
In other words, it is the effect of the tidal field on the mapping from Lagrangian to Eulerian positions.

The most obvious observational consequence of the effect described in \refeq{d2tintro} is an \textit{anisotropic distortion of the local small-scale power spectrum}, whose fractional amplitude is given by
\ba
& \frac{P(\vk_S,\tau|h_{ij}^{(0)})}{P_L(k_S, \tau)} - 1 \label{eq:Pkintro}\\
&\quad = 2 \hat k_S^i \hat k_S^j h_{ij}^{(0)} 
\left[ \alpha(k_L,\tau) - \beta(k_L,\tau) \frac{d\ln P_L(k_S,\tau)}{d\ln k_S}\right]\,.
\nonumber
\ea
Given that we consider the regime $k_L/k_S \ll 1$, this expression can
then be used to derive the contribution to the tensor-scalar-scalar bispectrum
$\< h_{ij}^{(0)}(\vk_L) \d(\vk_S) \d(\vk_S) \>$ in the squeezed limit.  Of
course, this bispectrum is only accessible observationally if one has an independent measurement of the vector or tensor modes. Alternatively, if no indepedent estimate
of $h_{ij}^{(0)}$ is available, \refeq{Pkintro} also describes a specific,
anisotropic contribution to the collapsed limit of the four-point function
of the density field \cite{jeong/kamionkowski:2012}, which can be measured without any external data sets.  The specific case studied by
\cite{jeong/kamionkowski:2012} was the 21cm emission at high redshifts.  
However, the same idea, applied to long-wavelength scalar tidal fields,
also underlies the tidal field reconstruction of \cite{pen/etal:12}.  

Another probe of anisotropic small-scale fluctuations is the
\textit{alignment of dark matter halos}, that are known to orient along 
long-wavelength tidal fields, and that in turn influence the
orientations of galaxies within them \cite{Peebles:69,CatelanKamionkowskiBlandford,HirataSeljak:04}.  This can be observed through
the shapes of galaxy images, as measured in large area weak lensing shear
surveys.  The preferential alignment of galaxy images with large-scale
tidal fields is known as \emph{intrinsic alignments}.  \citet{GWshear}
were the first to point out that tensor modes are also expected to
contribute to intrinsic alignments.  As shown there, the observed galaxy shape 
correlations induced by tensor modes are in fact expected
to be dominated by the alignment contribution.  The reason for this is
that the tensor-mode lensing effect (part of the projection effects
discussed above) is strongly suppressed as the propagating
gravitational waves fail to produce a coherent deflection
along the light cone as scalar perturbations do.  

However, the 
estimates of the alignment effect in \cite{GWshear} did not take into account
the qualitatively different evolution with time of tensor with respect
to scalar tidal fields.  Specifically, they assumed that the 
alignment is proportional to the instantaneous tidal field at the time
of observation of the galaxy.  On the other hand, here we found,
in agreement with \cite{dai/etal:13}, that the tidal effect on small-scale
fluctuations comes instead from a time integral over the past history of the tensor mode that peaks at horizon crossing $k_{L}\simeq \cH$ but remains constant afterwards.  In \refsec{IA} of this paper, we use these results to provide a more
accurate estimate of the intrinsic alignment by tensor modes. 
We find a shear B-mode power spectrum of $l(l+1)C^{BB}_{\gamma}(l)/2\pi \sim \mathrm{few} \times 10^{-12}$ over a wide range of scales and redshifts (see \reffig{Cl}).

The remainder of the paper derives the coefficient functions
$\alpha$ and $\beta$, and presents these applications in more detail.  
The outline of the paper is as follows. In \refsec{FNC} we review Fermi Normal Coordinates (\fnc) and their conformal analog (\fncb) introduced in \cite{conformalfermi} (leaving details to \refapp{FNC}) and derive a general expression for the local tidal field. In \refsec{LPT} we solve for the effect of the tidal field on short scale density fluctuations using Lagrangian Perturbation Theory (the equivalent Eulerian derivation can be found in \refapp{eulerian}). We give explicit applications of our general results to long-wavelength scalar, vector and tensor perturbations in \refsec{scalar} and \refsec{tensor}.  We discuss projection effects and the distortion of small-scale correlations observed on Earth
in \refsec{xilocal}.  Finally, in \refsec{IA} we derive the implications of our results for lensing shear surveys. 

Our notation is summarized in \reftab{not}.  
For our numerical results, we adopt a flat $\Lambda$CDM cosmology with
$\Omn = 1-\Omega_{\Lambda 0} = 0.3$ and $h=0.72$.

%%%%%%%%%%%%%%%%%%%%%%%%%%%%%%%%%%%%%%%%%%%%%%%%%%%%%%%%%%%%%%%%%%%%%%%%%%%

\section{Fermi Normal Coordinates}
\label{sec:FNC}

Our framework for the computation of gravitational tidal effects is the
conformal Fermi normal coordinate (\fncb) frame, which was first introduced
in \citet{conformalfermi} and which we review in this section.  

Consider a perturbed FRW metric given by
\be
g_{\mu\nu}(\tau,\vx)=a(\tau)^{2}\left[ \eta_{\mu\nu}+h_{\mu\nu}(\tau,\vx) \right]\,,
\label{eq:metrichij}
\ee
with $\eta_{\mu\nu}$ the mostly positive Minkowski-space metric and $h_{\mu\nu}\ll 1$ some set of small perturbations. We want now to study the effect of the interactions between long ($k_{L}$) and short ($k_{S}$) wavelength perturbations in $h$, assuming $k_{L}\ll k_{S}$. For this purpose we consider a region around the timelike geodesic of a comoving observer governed by the metric \refeq{metrichij}, with approximate size $k_{S}^{-1}$ on a certain spatial slice around the geodesic.  We can then construct a coordinate frame $\{\bar x_F^\mu\}$ with spatial origin corresponding to this central geodesic in which the metric is Friedmann-Robertson-Walker along the central geodesic, with corrections going as the spatial distance from the geodesic squared (the explicit form is given in \cite{conformalfermi}, see also \refapp{FNC}) \emph{at all times} $\bar x_F^0$
\be
g_{\mu\nu}^F = a^2(x^0[\bar x_F^0]) \left[\eta_{\mu\nu} + \O([\bar x_F^i]^2) \right]\,.
\label{eq:gmnF}
\ee
Note that these coordinates are not globally valid, but apply in a ``spaghetti-shaped''
region of spacetime around the central geodesic.  We call the frame described 
by the coordinates $\{\bar x_F^\mu\}$ the \emph{conformal} Fermi
Normal Coordinate frame (\fncb), as it
is a generalization of the Fermi Normal Coordinates (\fnc) first introduced
by \cite{ManasseMisner} (and recently applied in cosmology
in \cite{BaldaufEtal,stdruler,GWshear}).  
We give here a brief overview of its properties and refer the reader to \cite{conformalfermi} for further details.  When the size $\sim k_S^{-1}$ of the region considered
is much smaller than the horizon, then we can do a further simple coordinate
transformation to recover the standard FNC.  However, unlike the standard FNC,
\fncb~are also applicable if the region considered is superhorizon.  

In case of the standard FNC, the $\O(x_F^2)$ corrections are given by the 
Riemann tensor evaluated along the central geodesic.  For the metric 
\refeq{metrichij} (at linear order in $h$), this includes
terms of order $H^2$, $H \nabla h$, and $\nabla\nabla h$ (here $\nabla$ stands
for either a space or time derivative).  As we show 
in \refapp{FNC}, the $\O(\bar x_F^2)$ corrections in \fncb~on the other hand 
come from two sources:  first, there are contributions of order $\nabla\nabla h$
from the Riemann tensor of the \emph{conformal} metric $\eta_{\mu\nu}+h_{\mu\nu}$, which agree with the corresponding terms in \fnc~(up to factors of $a$ 
from the leading order relation $x_F^i = a\,\bar x_F^i$ between spatial \fnc~and \fncb).  
Second, there are terms of order $H \nabla h$ which enter through the $a^2$ 
prefactor due to the transformation of the time coordinate.  Again,
these agree with the terms of the same type in \fnc.  Finally, the 
$\O(H^2 x_F^2)$ terms disappear in \fncb, since we have explicitly kept the
$a^2$ prefactor.  

Thus, the \fncb~correspond to the natural \emph{comoving} coordinates an
observer moving along the central geodesic would choose.  In fact, 
the coordinates chosen to interpret cosmological observations from Earth
are essentially \fncb~constructed for the geodesic of the Solar System
and with the size of the patch given by our current Hubble horizon, thereby 
removing effects of all super-horizon modes $k_L < H_0$, where again $k_L$ is the 
\emph{comoving} wavenumber of the long-wavelength perturbation.  
As mentioned above, the advantage of the \fncb~frame over \fnc~is that the 
corrections to the FRW metric are always of order $(\bar x_F k_L)^2$
[so in case of the \fncb~around Earth, the parameter is $(k_L H_0)^2$].  
This allows one to follow the given region around the central geodesic 
back to early times where it was larger than the horizon $H^{-1}$ (at which 
point the regular FNC become invalid).  The \fncb~frame is useful for
studying the gravitational interaction of perturbations in cosmology whenever
there is a hierarchy between long and short modes, so that the effect of 
the long modes can be studied neglecting corrections of higher order in $k_L/k_S$.  

%%%%%%%%%%%%%%%%%%%%%%%%%%%%%%%%%%%%%%%%%%%%%%%%%%%%%%%%%%%%%%%%%%%%%%%%%%%%
\subsection{Non-relativistic limit}
\label{sec:FNCA}

Note that since we consider scalar\footnote{The framework of the previous section is valid to study any type of small scale perturbations but in this paper we will focus on short \textit{scalar} perturbations.} perturbations
that are much smaller in scale than tensor perturbations $k_{S}\gg k_{L}$, and 
since the effect of the latter only becomes relevant as $k_L\gtrsim \cH$, for all practical purposes we can restrict our analysis to when the short scalar fluctuations are well within the horizon, $k_{S}\gg \cH$.  
Additionally, in this work we will restrict ourselves to the dynamics of non-relativistic matter.  These two assumptions allow us to use the standard pseudo-Newtonian limit usually adopted in the theory of large-scale structure to describe the gravitational dynamics on short scales. 

Throughout, we neglect the effect of perturbations in the radiation component,
which is not correct in general: radiation interacts gravitationally with matter at all times. Also, before recombination, radiation couples tightly to baryons. There are two regimes in which we can neglect the effect of radiation on matter. The first is during matter domination, i.e. for $a \gg a_{\rm eq} = \Omega_{r0}/\Omn$ (and hence after recombination) when the energy density of radiation has redshifted away and its gravitational coupling with matter is very small. The second is during radiation domination on scales smaller than the dissipation scale, where the perturbations in the electron-baryon-photon fluid are completely erased.  Hence our results will be applicable in these two regimes. On the other hand, the scales that enter during radiation domination but are larger than the dissipation scale are quite interesting observationally, and deserve a separate study with a full relativistic treatment. We defer this to future work.

Consider objects in the vicinity of the central geodesic around which 
the \fncb~are constructed.  If these objects are slow moving, i.e.~if their
velocities relative to that of the central geodesic are much smaller than
the speed of light, then their dynamics are governed up to order $v/c$ 
by $g_{00}^F$.  In the following, we will assume that in global
coordinates $h_{0i}=0$, which applies to vector and scalar perturbations in
popular gauges as well as tensor perturbations.  
In this case, $g_{00}^F$ is given by (see \refapp{FNC})
\ba
g_{00}^F =\:& -a_F^2(\ebf) \Big[ 1 +2\Phi_s(\vx,\ebf) -\, t_{ij}(\ebf) \bar x_F^i \bar x_F^j\Big] \qquad
\label{eq:FNCmetric}
\\
%%%
t_{ij}(\ebf) =\:& 
 \frac12 \left[ a^{-1} (a h'_{ij})' + h_{00,ij}\right]\,.
\label{eq:tij}
\ea
Here, and throughout, primed denote derivatives with respect to $\tau$.  
In \refeq{tij}, all occurences of $h$ are evaluated at a given proper time $\ebf$ 
along the central geodesic (i.e. at $\bar\vx_F=0$ in \fncb).  $a_F$ is the
locally measured scale factor given by
\be
a_F(\ebf) \equiv a(\tau(\ebf,\v{0}))= a\left(\ebf + \frac12 \int^{\ebf} h_{00}(\v{0},\tau)d\tau\right)\,.
\ee
The apparent unphysical dependence on a metric perturbation $h_{00}$ (without any derivative) is simply because we are referring to an unobservable ``background'' scale factor $a(\tau)$ here.  
As discussed in more detail in \refapp{FNC}, this corresponds to an unobservable shift 
in the time coordinate.  What \emph{is}
observable is the different proper time $\ebf$ of different regions on a
constant-observed-redshift surface.  This is part of the ``projection effects''
we will discuss in \refsec{proj}.  

There are two contributions to $g_{00}^F$: the first, $\Phi_s$ is the potential sourced by the small-scale scalar perturbations.  The second, $t_{ij}$, is the tidal tensor induced by the long-wavelength metric perturbations.  Note that $t_{ij}$ has dimension 1/length$^2$.  We will work to linear order in $t_{ij}$ throughout.  Note that we allow for a non-zero trace component $t_i^{\,  i}$, which
will permit us to consider a long-wavelength density perturbation in addition
to vectors and tensors.  There are corrections to \refeq{FNCmetric} of order $(\bar x_F^i)^3$
which we neglect as they are suppressed by $k_L/k_S$.  

The effect of tensor and, for common gauges such as Newtonian and synchronous gauges, that of vector modes is encoded in
the $h_{ij}$ contribution to $t_{ij}$.  This contribution agrees with
that derived in \cite{GWshear}, taking into account that the tidal
field in the latter paper is given in terms of physical coordinates
$\vr_F = a \bar\vx_F$.  We see that in order to have an
effect, these modes have to evolve in time; specifically, superhorizon modes
which are conserved will have no effect.  On the other hand, scalar 
long-wavelength perturbations are encoded by the $h_{00} = -2 \Phi_L$
contribution, where $\Phi_L$ is the long-wavelength potential perturbation
in Newtonian gauge.  Note that the tidal effect is proportional to the
second \emph{spatial} derivative of $h_{00}$: this reflects the fact that spatially
constant and pure-gradient potential perturbations cannot have an observable 
impact by way of the equivalence principle.

In the following, we will work exclusively in the \fncb~frame.  Therefore
we simplify the notation in what follows:
\ba
\bar x_F^0 \to\:& \tau \vs
\bar x_F^i \to\:& x^i \vs
a_F \to\:& a\,.
\ea
By separating the linearized Einstein equations into long- and short-wavelength
parts and transforming the short-wavelength part to \fncb, once can show 
that up to corrections of order $k_L/k_S$,
the small-scale potential $\Phi_s$ satisfies 
the standard Poisson equation in comoving coordinates,
\be
\nabla^2 \Phi_s = 4\pi G a^2 \d\rho = \frac32 \Omn H_0^2 a^{-1} \d = \frac32 \Om(\tau) \cH^2 \d\,.
\label{eq:poisson}
\ee
where $G$ is Newton's constant, $\cH = a'/a$, $\rho$ is the matter density with homogenous average $\bar \rho$ and contrast $\delta\equiv \delta \rho/\bar \rho-1$. 

We define a transfer function $T(k_L,\tau)$ of the tidal field as follows,
\be
t_{ij}(\v{0};\tau) = T(k_L,\tau)\, t_{ij}^{(0)}(\v{0})\,,
\label{eq:transfer}
\ee
which in general depends on the 
wavenumber $k_L$ of the long-wavelength perturbation. We will give
the transfer function explicitly later on when dealing with the scalar
and tensor tidal fields separately.
In the following,
we will often suppress the argument $k_L$ as it does not enter in the
derivation otherwise.  
Throughout, we will assume that the source term for the second order
density goes to zero at early times, i.e. 
$a(\tau) t_{ij}(\tau) \stackrel{\tau\to 0}{=} 0$;  more specifically, we
assume that the small-scale fluctuations have settled into 
the growing mode by the time $a(\tau) t_{ij}(\tau)$ becomes non-negligible.  
This again is only valid if the wavelength of the small-scale fluctuations
is sufficiently smaller than that of the tidal field.

%%%%%%%%%%%%%%%%%%%%%%%%%%%%%%%%%%%%%%%%%%%%%%%%%%%%%%%%%%%%%%%%%%%%%%%%%%%
%%%%%%%%%%%%%%%%%%%%%%%%%%%%%%%%%%%%%%%%%%%%%%%%%%%%%%%%%%%%%%%%%%%%%%%%%%%
\section{Lagrangian derivation of tidal effects}
\label{sec:LPT}

In this section, we derive the effects of the external field on small-scale
density perturbations using the Lagrangian approach.  A Eulerian derivation
which arrives at the same result is given in \refapp{eulerian}.  
For simplicity, we will assume an Einstein-de Sitter Universe in this section.  
This is applicable to external tidal fields which become
relevant during matter domination, e.g. those induced by
long-wavelength modes that enter the horizon during matter domination, i.e.
with wave numbers $H_0^{-1} \ll k_L < k_{\rm eq}$.  
We extend the derivation to include tidal fields that become relevant during 
radiation domination and $\Lambda$ domination in \refsec{tensorRD}.

The comoving Eulerian coordinate $\vx$ at conformal time $\tau$ is related to
the Lagrangian coordinate $\v{q}$ by the displacement $\v{s}$,
\be
\vx(\v{q},\tau) = \v{q} + \v{s}(\v{q},\tau)\,.
\ee
In Lagrangian perturbation theory (LPT), we adopt the single-stream approximation,
in which case the Eulerian fractional matter overdensity $\d(\vx,\tau)$ is given by
\be
\d[\vx(\v{q},\tau),\tau] = |\v{1} + \v{M}(\v{q},\tau)|^{-1} - 1\,,
\label{eq:dLPT}
\ee
where $\v{M}_{ij}$ is the deformation tensor,
\be
\v{M}_{ij}(\v{q},\tau) = \frac{\partial}{\partial q^i} s^j(\v{q},\tau)\,.
\ee
Note that $\partial_q^i = \partial_x^i + \v{M}^i_{\  j} \partial_q^j$.  
The evolution equation for $\v{s}$ is simply the equation of motion of a particle in comoving units,
\ba
s''^i(\vq,\tau) + \cH s'^i(\vq,\tau) =\:& - \partial_x^i \left[\Phi_s(\vx) + \frac12 t_{kl} x^k x^l \right]_{\vx(\vq),\tau}
\,,
\label{eq:seom}
\ea
where primes denote derivatives with respect to conformal time $\tau$.  
Using \refeq{dLPT}, the Poisson equation \refeq{poisson} becomes at second order
\ba
& \nabla_x^2\Phi_s(\vx(\vq)) = \frac32 \Om \cH^2 \d(\vx(\vq)) \vs
& = \frac32 \Om \cH^2
\bigg[-\Tr \v{M} + \frac12\left( (\Tr \v{M})^2 + \Tr(\v{M}\cdot\v{M})\right) 
\vs
& \hspace*{2.5cm} + \O( (\v{M})^3)\bigg]_{\vq}\,.
\label{eq:Phis}
\ea
We are interested in the leading effect of $t_{ij}$ on the density
in Eulerian space \refeq{dLPT}.  For this, we decompose the displacement
as
\be
\v{s} = \v{s}_s + \v{s}_t\,,
\ee
where $\v{s}_s$ is the scalar contribution which remains when setting $t_{ij}$ to zero.  
Correspondingly, we will use $\v{M}_s,\,\v{M}_t$.  
We will further perform a perturbative expansion in $\v{s}$.  Specifically,
we consider the linear displacement, which uniquely separates into scalar
and tensor pieces $\v{s}_{1,s},\,\v{s}_{1,t}$, and the quadratic mixed
contribution from the coupling of $\v{s}_s$ and $\v{s}_t$, denoted as
$\v{s}_{2,t}$ (the contributions of order $(s_{1,s})^2$ lead to the standard
second order LPT result).  Without loss of generality, we set $\v{s}_{1,s}(\vq=0,\tau) = \v{0} = \v{s}_{1,t}(\vq=0,\tau)$ at some time of interest $\tau$, so that 
\emph{at linear order} the origin 
coincides in both Eulerian and Lagrangian coordinates.

% % % % % % % % % % % % % % % % % % % % % % % % % % % % % % % % % % % % % %
\subsection{Linear solutions}

The linearized version of \refeq{seom} becomes, separated into scalar and tensor parts,
\ba
s_{1,s}''^i(\vq,\tau) + \cH s_{1,s}'^i(\vq,\tau) =\:&
\frac32 \cH^2 \frac{\partial_q^i}{\nabla_q^2} \partial_{q\,j} s_{1,s}^j(\vq,\tau) \vs
s_{1,t}''^i(\vq,\tau) + \cH s_{1,t}'^i(\vq,\tau) =\:&
- \frac12 \partial_q^i \left[ t_{kl}(\tau) q^k q^l \right]\,.
\label{eq:seomlin}
\ea
Since this is at linear order, we have set $\vx = \vq$.  
Assuming only the growing mode is present in the initial conditions,
the first equation can be integrated to give
\be
s_{1,s}^i(\vq,\tau) = -\frac{\partial_q^i}{\nabla_q^2} \d_{1,s}(\vq,\tau)
= - a(\tau) \frac{\partial_q^i}{\nabla_q^2} \d_{1,s}(\vq,\tau_0)\,,
\label{eq:sLs}
\ee
where $a(\tau_0)=1$.  The equation for $s_{1,t}^i$, rewritten as
\be
\left(\tau^2 s_{1,t}'^i\right)' = - \tau^2 T(\tau) t^{(0)\,i}_{\quad\    k} q^k \,,
\ee
can be integrated to give
\ba
s_{1,t}^i(\vq,\tau) =\:& - F(\tau) t^{(0)\,i}_{\quad\    k} q^k 
\label{eq:sLt}\\
F(\tau) \equiv\:& \int_0^\tau \frac{d\tau'}{a(\tau')} \int_0^{\tau'} d\tau'' a(\tau'') T(\tau'')\,.
\label{eq:Fdef}
\ea
Note that
\be
T(\tau) = \frac1{a(\tau)} \left[a\,F'(\tau)\right]'\,,
\ee
and that \refeqs{sLt}{Fdef} are valid for a general expansion history.  In parallel with
the linear scalar density $\d_{1,s}$, we define
\be
\d_{1,t}(\vq,\tau) = - \partial_{q\,i} s^i_{1,t}(\tau) = F(\tau) t^{(0)\,i}_{\quad\   i}\,.
\label{eq:dLt}
\ee
If $t_k^{\  k}=0$, ${\rm Tr}\,\v{M}_{1,t} = 0$ and
there is no first-order contribution to the density ($\d_{1,t}=0$) as
expected.

% % % % % % % % % % % % % % % % % % % % % % % % % % % % % % % % % % % % % %
\subsection{Second-order solution}

The equation for the second-order displacement $s_{2,t}$ is obtained by collecting all second-order pieces from the right-hand side of \refeq{seom}.  
As described above, we will only consider the coupling of $s_{1,s}$ with
$s_{1,t}$.  

Taking the divergence with respect to $\vq$ of the equation for $\v{s}$ 
[\refeq{seom}] yields
\ba
\s'' + \cH \s' =\:& -\nabla_x^2 \Phi_s - \v{M}_{ij} \partial_x^i \partial_x^j \Phi_s \vs
& - \v{M}_{ij} \partial_x^i \partial_x^j \left[ \frac12 t_{kl} x^k x^l \right]\,,
\label{eq:sigeom}
\ea
where $\s = \partial_{q\,i} s^i$ and the r.h.s. is evaluated at $\vx(\vq)$.  Inserting the Poisson equation
at second order [\refeq{Phis}], writing $\v{M} = \v{M}_s + \v{M}_t$, and
subtracting the equation for $\v{s}_{1,t}$ [\refeq{seomlin}] 
leads to the following equation for $\s_{2,t}$:
\ba
\s_{2,t}'' + \cH \s_{2,t}' =\:& \frac32 \Om \cH^2 \Tr \v{M}_{2,t} 
\label{eq:s2Heom}\\
& -\frac32 \Om \cH^2 \Tr \v{M}_{1,s} \Tr \v{M}_{1,t} - \v{M}_{1,s}^{ij} t_{ij}\,,
\nonumber
\ea
where on the r.h.s. all contributions are evaluated at $\vq$.  
Note that the second term in \refeq{sigeom} has canceled with the term
$\propto \Tr \v{M}_s \cdot \v{M}_t$ from the second-order density.  
We obtain
\ba
& \s_{2,t}'' + \cH \s_{2,t}' - \frac32 \Om \cH^2 \s_{2,t} \vs
%
%& = -\frac32 \Om \cH^2 \Tr \v{M}_{1,s} \Tr \v{M}_{1,t} -\v{M}_{1,s}^{ij} t_{ij}(\tau) \vs
%
& = -\frac32 \Om \cH^2 \d_{1,s} \d_{1,t}\Big|_{\vq,\tau} + \left(\frac{\partial^i\partial^j}{\nabla^2} \d_{1,s}\right)_{\vq,\tau} t_{ij}(\tau)\,.
\label{eq:s2t1}
\ea
Specializing to Einstein-de Sitter, the previous equation becomes
\ba
& \s_{2,t}''(\vq,\tau) + \frac2{\tau} \s_{2,t}'(\vq,\tau) - \frac6{\tau^2}\s_{2,t}(\vq,\tau) = \Sigma(\vq,\tau) \vs
& \Sigma(\vq,\tau) = - \frac32 H_0^2 a(\tau) \d_{1,s}(\vq,\tau_0) F(\tau) t^{(0)\,i}_{\quad\  i}  \vs
& \hspace*{0.5cm}
+ a(\tau) T(\tau) \left(\frac{\partial^i\partial^j}{\nabla^2} \d_{1,s}(\vq,\tau_0)\right) t_{ij}^{(0)}
\,.
\ea
The growing and decaying modes of this equation correspond to 
$\s_{2,t} \propto \tau^2$ and $\s_{2,t} \propto \tau^{-3}$, respectively,  
and the solution is
\be
\s_{2,t}(\vq,\tau) = \int_0^\tau d\tau' \frac15\left[\frac{\tau^2}{\tau'} - \frac{\tau'^4}{\tau^3}\right] \Sigma(\vq,\tau')\,.
\ee
By assumption (see the last paragraph of \refsec{FNCA}), 
$\Sigma(\vq,\tau) \to 0$ as $\tau\to 0$, and we have fixed the 
boundary conditions so that both $\s_{2,t}$ and  $\s_{2,t}'$ vanish in this limit as well.  
We then obtain
\ba
\s_{2,t}(\vq,\tau) =\:& D_{\s1}(\tau) \left(\frac{\partial^i\partial^j}{\nabla^2} \d_{1,s}(\vq,\tau)\right) t_{ij}^{(0)} \vs
& - \frac32 D_{\s2}(\tau) \d_{1,s}(\vq,\tau) t_{\  i}^{(0)\,i}\,,
\ea
where the coefficient functions are given by
\ba
V(\tau) \equiv\:& \int_0^{\tau} d\tau' \frac{\tau'^5}{\tau^5} F'(\tau') 
\label{eq:Vdef}\\
%%%
D_{\s1}(\tau) \equiv\:&  \frac1{a(\tau)}\int_0^\tau d\tau' \frac15\left[\frac{\tau^2}{\tau'} - \frac{\tau'^4}{\tau^3}\right] a(\tau') T(\tau') \vs
=\:& \frac15 \left\{ F(\tau) + 4 V(\tau) \right\}
\label{eq:Dsigma1}\\
%%%
D_{\s2}(\tau) \equiv\:&  \frac{H_0^2}{a(\tau)} \int_0^\tau d\tau' \frac15\left[\frac{\tau^2}{\tau'} - \frac{\tau'^4}{\tau^3}\right] F(\tau') 
\,.
\label{eq:Dsigma2}
\ea
Let us now derive the \emph{Eulerian} density in the presence of $t_{ij}$.  
This is defined as
\be
\d_{t}(\vx) = \d(\vx) - \d(\vx)|_{t_{ij}=0}\,.
\label{eq:dHdef}
\ee
Using \refeq{dLPT}, we have
\ba
& \d_t(\vx(\vq)) = \d_{1,t}(\vq) + \d_{2,t}(\vx(\vq)) \vs
& = \d_{1,t}(\vq) - \s_{2,t}(\vq) + \d_{1,t}(\vq) \d_{1,s}(\vq) + \Tr\left(\v{M}_{1,s}\cdot \v{M}_{1,t}\right)_{\vq}\,.\nonumber
\ea
However, there is a further subtlety in \refeq{dHdef}: we want to compare
the overdensities at the same \emph{Eulerian} position $\vx$.  The \emph{Lagrangian}
coordinate that corresponds to this position is different for $t_{ij} \neq 0$
and $t_{ij}=0$.  
More precisely, we have (at linear order which suffices for this purpose)
\be
\vx = \vq + \v{s}_{1,s} + \v{s}_{1,t}\,,
\ee
whereas $\d_{1,s}$ as defined here gives the density at
\be
\vx_s = \vq + \v{s}_{1,s} = \vx - \v{s}_{1,t}\,.
\ee
Thus, the contribution to the Eulerian density by tensor modes is 
\ba
\d_{t}(\vx,\tau) =\;& \d_{1,t}(\vx) - \s_{2,t}(\vx) + \d_{1,t}(\vx) \d_{1,s}(\vx) \vs
& + \Tr\left(\v{M}_{1,s}\cdot \v{M}_{1,t}\right)_{\vx} - (\vx-\vx_s)\cdot\vn_\vq \d_{1,s}(\vq)\Big|_{\vx} \vs
%%%
=\:& \d_{1,t}(\vx) - \s_{2,t}(\vx) + \d_{1,t}(\vx) \d_{1,s}(\vx) \vs
& + \Tr\left(\v{M}_{1,s}\cdot \v{M}_{1,t}\right)_{\vx} - s_{1,t}^i(\vx) \partial_{q\,i} \d_{1,s}(\vx)\,.
\label{eq:dtLPT1}
\ea
Note that we can replace $\vq$ with $\vx$ at this order.  
Inserting the expressions derived above, we obtain for the second order contribution
\ba
\d_{2,t}(\vx,\tau) =\:& t_{ij}^{(0)}\bigg[
- D_{\s1}(\tau) \frac{\partial^i\partial^j}{\nabla^2}  
+ \frac32 \d^{ij} D_{\s2}(\tau) \vs
&\quad+ F(\tau) \left\{\d^{ij} + \frac{\partial^i \partial^j}{\nabla^2}
+  x^i \partial^j\right\}
\bigg] \d_{1,s}(\vx, \tau)\,.
\label{eq:d2tLPT1}
\ea
To recap, the first two terms here come from the modified second-order 
evolution of the scalar fluctuations, due to the presence of the external
tidal field proper (first term) and the additional contribution to the
matter density (second term).  The next two terms are due to the non-linear
relation between displacement and density, which entails a coupling
of the linear displacements due to small-scale scalar and external
tidal displacements.  Finally, the last term is directly proportional to
the displacement by the external tidal field, and encodes the fact that
we evaluate the small-scale perturbations at different relative (Eulerian) 
positions than we would have in the absence of $t_{ij}$.  

We can now bring \refeq{d2tLPT1} into the form of \refeq{d2tintro} in
\refsec{intro}:
\ba
& \d_{2,t}(\vx,\tau) \vs
& \quad = t_{ij}^{(0)}(\vx) \left[ \alpha(\tau) \frac{\partial^i\partial^j}{\nabla^2} + \beta(\tau) x^i \partial^j + \gamma(\tau) \d^{ij} \right] \d_{1,s}(\vx,\tau)\,,
\label{eq:d2tLPT}
\ea
where
\ba
\alpha(\tau) =\:& \frac45 \left\{ F(\tau) - V(\tau) \right\} \vs
\beta(\tau) =\:& F(\tau) \vs
\vs
\gamma(\tau) =\:& \frac32 D_{\s2}(\tau) + F(\tau) 
\,.\label{eq:d2tLPT2}
\ea
Again, there is a dependence on $k_L$ which we have not written here
that enters through
the tranfer function in \refeq{Fdef} and \refeqs{Vdef}{Dsigma2}.

%%%%%%%%%%%%%%%%%%%%%%%%%%%%%%%%%%%%%%%%%%%%%%%%%%%%%%%%%%%%%%%%%%%%%%%%%%%%%%%%
%%%%%%%%%%%%%%%%%%%%%%%%%%%%%%%%%%%%%%%%%%%%%%%%%%%%%%%%%%%%%%%%%%%%%%%%%%%%%%%%
\section{Scalar tidal field}\label{sec:scalar}

In this section, we consider a tidal field $t_{ij}$ induced by a long-wavelength
density perturbation $\d_{1,L}$.  In principle we could perform this computation for any $k_{L}\ll k_{S}$. On the other hand, our goal is to make contact with standard results, thus providing a non-trivial check of our formulae. Therefore we will assume to be in matter domination and that the long mode is well inside the horizon, namely $k_L \gg \cH$.  In this case, \refeq{d2tLPT} should reduce to the standard expression ($F_{2}$ kernel of \cite{bernardeau/etal:02}) for the second order density perturbation in the limit that one mode is much longer than the other.

Before discussing the detailed calculation it is useful to present a heuristic but intuitive derivation of the main result of this section. We want to predict the structure of the second order density field $\delta_{2}$. Up to a potentially time-dependent factor, this is the same as the quadratic source terms in the Eulerian equation of motion for $\delta$, \refeq{d2eomcomp3}. To avoid proliferation of $\nabla^{-2}$ it is convenient to use the Newtonian potential and its derivatives as fundamental building block $\Phi\sim \left( \cH^{2}/k^{2} \right) \delta$, instead of $\delta$ itself. Then, we should write all possible second order terms using just $\Phi$ and spatial derivatives, since time derivatives are very small during matter domination (zero in exact EdS). First we notice that $\Phi$ must appear with at least one derivative (a constant $\Phi$ cannot lead to any physical effect); the term with one derivative contains a locally-unobservable uniform acceleration or bulk flow as we will discuss below. Second, since $\d$ is a scalar we need to have an even total number of derivatives (lest we are left with uncontracted indices). The lowest number of derivatives is then two but the term $\partial_{i}\Phi \partial_{i}\Phi $ cannot appear because the equation for $\delta$ has an additional spatial derivative with respect to the Euler equation (conservation of momentum in \refeq{euler}) where each $\Phi$ should appear with at least one spatial derivative. Hence the allowed second order terms have at least four derivatives:
\be
\partial_{i}\partial_{j}\Phi\partial_{i}\partial_{j}\Phi\,,\,\partial^{2}\Phi\partial^{2}\Phi,\  \mathrm{and}\   \partial_{i}\Phi \partial_{i}\partial^{2}\Phi\,.\label{eq:vici}
\ee 
Terms with a higher number of spatial derivatives are suppressed by $k/k_{\rm NL}$ where $k_{\rm NL}$ is the cutoff of the hydrodynamic theory or fluid approximation (see e.g.~\cite{Baumann:2010tm}) and we can safely neglect here as long as we are interested in the mildly (as opposed to fully) non-linear regime. The terms in \refeq{vici} are indeed those appearing in the well-known $F_2$ kernel.  We can further massage these terms by applying to the case at hand. We take one perturbation to be long and one short, dropping terms suppressed by $k_{L}/k_{S}$. We are then left with only three terms:
\be\label{eq:guess1}
\partial_{i}\partial_{j}\Phi_{L}\partial_{i}\partial_{j}\Phi_{S}\,,\,\partial^{2}\Phi_{L}\partial^{2}\Phi_{S},\  \mathrm{and}\  \partial_{i}\Phi_{L} \partial_{i}\partial^{2}\Phi_{S}\,.
\ee 
The presence of $\partial_{i}\Phi_{L}$ in an observable quantity like $\delta_{2}$ tells us immediately that this expression is valid in some set of global coordinates that are \textit{not} free falling. In order to derive the second order effects that a local free falling observer would measure, we expand $\partial_{i}\Phi_{L}$ around the observer's geodesic
\be\label{eq:guess2}
\partial_{i}\Phi_{L}(\v{x})=\partial_{i}\Phi_{L}(\v{0})+\partial_{i}\partial_{j}\Phi_{L}(\v{0})x^{j}+\dots
\ee
The first term in this expression, once contracted with $\partial_{i}\partial^{2}\Phi_{S}$, represents the effect of the bulk flow since it can be thought of as arising from expanding $\partial^{2}\Phi_{S}(x+U_{L} t)$ at linear order in $u_{L}$ and using $u_{L}^{i}\sim \partial^{i}\Phi_{L}/\cH$. This effect is present in global coordinates [as used in standard perturbation theory, see \refeq{d2tscalar} below] but can be removed by a boost into the free falling local frame where $\partial_{i}\Phi$ vanishes.

Now that we have built some intuition, let us move to the detailed derivation. Using \refeq{tij} and the fact that in global Newtonian coordinates
\be
h_{00}=-2\Phi_{L}\,,\quad h_{ij}=-2\Phi_{L}\delta_{ij}\,,
\ee
we find
\be
t_{ij}=- \left[  \Phi_{L,ij}+\delta_{ij} \left(  \cH \Phi_{L}'+\Phi_{L}''\right)\right]\simeq -\Phi_{L,ij}\,,
\ee
since $\Phi$ is constant during matter domination.  
The transfer function [\refeq{transfer}] is then $T(k_L,\tau) = $~const
and we can choose for simplicity $T(\tau,k_{L})=1$.  Then one finds
\ba
F(\tau) =\:& \frac23 H_0^{-2} a(\tau)
\vs
%%%
V(\tau) =\:& D_{\s2}(\tau)
= \frac4{21} H_0^{-2} a(\tau)
\,.
\ea
Further, evaluating the Poisson equation at $\tau_0$ where $a(\tau_0)=1$,
we have
\be
t_{ij}^{(0)} = \frac32 H_0^2 \frac{\partial^i \partial^j}{\nabla^2} \d_{1,L}(\tau_0)\,.
\ee
\refeq{d2tLPT}
then yields
\ba
& \d_{2,s}(\vx,\tau) =
\frac47 \left(\frac{\partial^i \partial^j}{\nabla^2} \d_{1,L}(\tau)\right)
\left(\frac{\partial^i \partial^j}{\nabla^2} \d_{1,s}(\vx,\tau)\right) \vs
&+ \frac{10}7 \d_{1,L}(\tau) \d_{1,s}(\vx, \tau)
+ \left(\frac{\partial^i \partial^j}{\nabla^2} \d_{1,L}(\tau)\right)
x^i \partial^j \d_{1,s}(\vx, \tau)
\label{eq:d2tscalar1}
\,.
\ea
This expression contains precisely the terms we predicted in \refeq{guess1} and \refeq{guess2}, without the bulk flow [first term in \refeq{guess2}]. In order to compare with the standard perturbation theory result, we have to transform
from the local \fncb~(free falling) frame to global coordinates, which
corresponds to adding back in the bulk flow. We can hence replace
\be
\left(\frac{\partial_i\partial_j}{\nabla^2} \d_{1,L}\right) x^j \to \frac{\partial_i}{\nabla^2} \d_{1,L}\,.
\label{eq:d1L}
\ee
Further, note that the other permutation corresponding to this term, i.e.
$(\partial_i/\nabla^2 \d_{1,s})\partial_i \d_{1,L}$ is not included in our
derivation since it involves the third derivative of the long-wavelength potential, i.e. it is suppressed by $k_L/k_S$.  Finally, 
in our treatment we have split $\d_1 = \d_{1,L} + \d_{1,s}$ so that the 
second order solution contains two permutations $L\leftrightarrow s$ 
in the quadratic source terms.  Thus, if we express the result in terms of $\d_1$, we need to divide \refeq{d2tscalar1} by two.  We finally obtain
\ba
& \d_{2}(\vx,\tau) =
\frac27 \left(\frac{\partial_i \partial_j}{\nabla^2} \d_{1}(\vx,\tau)\right)
\left(\frac{\partial^i \partial^j}{\nabla^2} \d_{1}(\vx,\tau)\right) \vs
& + \frac{5}7 \d_{1}(\vx,\tau) \d_{1}(\vx, \tau) 
+ \left(\frac{\partial_i}{\nabla^2} \d_{1}(\vx,\tau)\right) \partial^i 
\d_{1}(\vx, \tau)
\label{eq:d2tscalar}
\ea
This is easily seen to be identical to the standard second order scalar density
perturbation, which is usally expressed in Fourier space through the $F_2$
kernel (e.g., \cite{bernardeau/etal:02}).

%%%%%%%%%%%%%%%%%%%%%%%%%%%%%%%%%%%%%%%%%%%%%%%%%%%%%%%%%%%%%%%%%%%%%%%%%%%
%%%%%%%%%%%%%%%%%%%%%%%%%%%%%%%%%%%%%%%%%%%%%%%%%%%%%%%%%%%%%%%%%%%%%%%%%%%
\section{Tensor and vector modes}
\label{sec:tensor}

In this section, we describe the effect of vector and tensor metric
perturbations on the growth of small-scale fluctuations.  We will focus 
mostly on tensor modes.  All models of
inflation at sufficiently high energy scale predict an approximately
scale-invariant background of gravitational waves, providing strong
motivation to study their effects.  On the other hand, it is very
difficult to devise a mechanism which produces long-wavelength vector 
modes while satisfying all cosmological constraints e.g. from the CMB.  
Our formalism is not suited to treat effects of
very small scale vector modes which could have been produced in the
early Universe, since we always consider scalar perturbations
with wavelengths much smaller than those of the vector modes 
(in any case, the effect of small scale vector modes is hard to observe
unless they are coupled to long-wavelength fluctuations).  However,
all of the following results immediately apply to vector modes
should the need arise; the only modification is in the transfer
function of the long-wavelength tidal field [\refeqs{Dhtensor}{Dhvector}].

%%%%%%%%%%%%%%%%%%%%%%%%%%%%%%%%%%%%%%%%%%%%%%%%%%%%%%%%%%%%%%%%%%%%%%%%%%%
\subsection{Tidal field}

Since we work to linear order in the long-wavelength perturbations, any 
scalar long-wavelength perturbations in $h_{\mu\nu}$ decouple and need not 
be considered.  We thus set $h_{00} = 0$ in \refeq{tij}, and
the long-wavelength tensor and vector modes are contained in $h_{ij}$.  

We define a general transfer function $D_h(\tau)$ so that
\be
h_{ij}(\v{0},\tau) = D_h(\tau) h_{ij}^{(0)} \,,
\ee
where $h_{ij}^{(0)} = h_{ij}(\v{0},\tau=0)$ is the primordial amplitude
of the metric perturbation and $D_h(\tau\to 0) = 1$.  
For a single Fourier mode $k_L$ tensor perturbation during matter 
domination, this transfer function is given by
\ba
D_{h,T}(\tau) =\:& 3 \frac{j_1(k_L\tau)}{k_L\tau} \,,
\label{eq:Dhtensor} 
\ea
while the decay of vector modes produced instantaneously at a time $\tau_*$ 
is described by
\be
D_{h,V}(\tau) = \frac{a^2(\tau_*)}{a^2(\tau)} \,,
\label{eq:Dhvector}
\ee
and $D_{h,V}(\tau) = 0 $ for $\tau < \tau_*$.

%!!!!!!!!!!!!!!!!!!!!!!!!!!!!!!!!!!!!!!!!!!!
\begin{figure}[t!]
\centering
\includegraphics[width=0.48\textwidth]{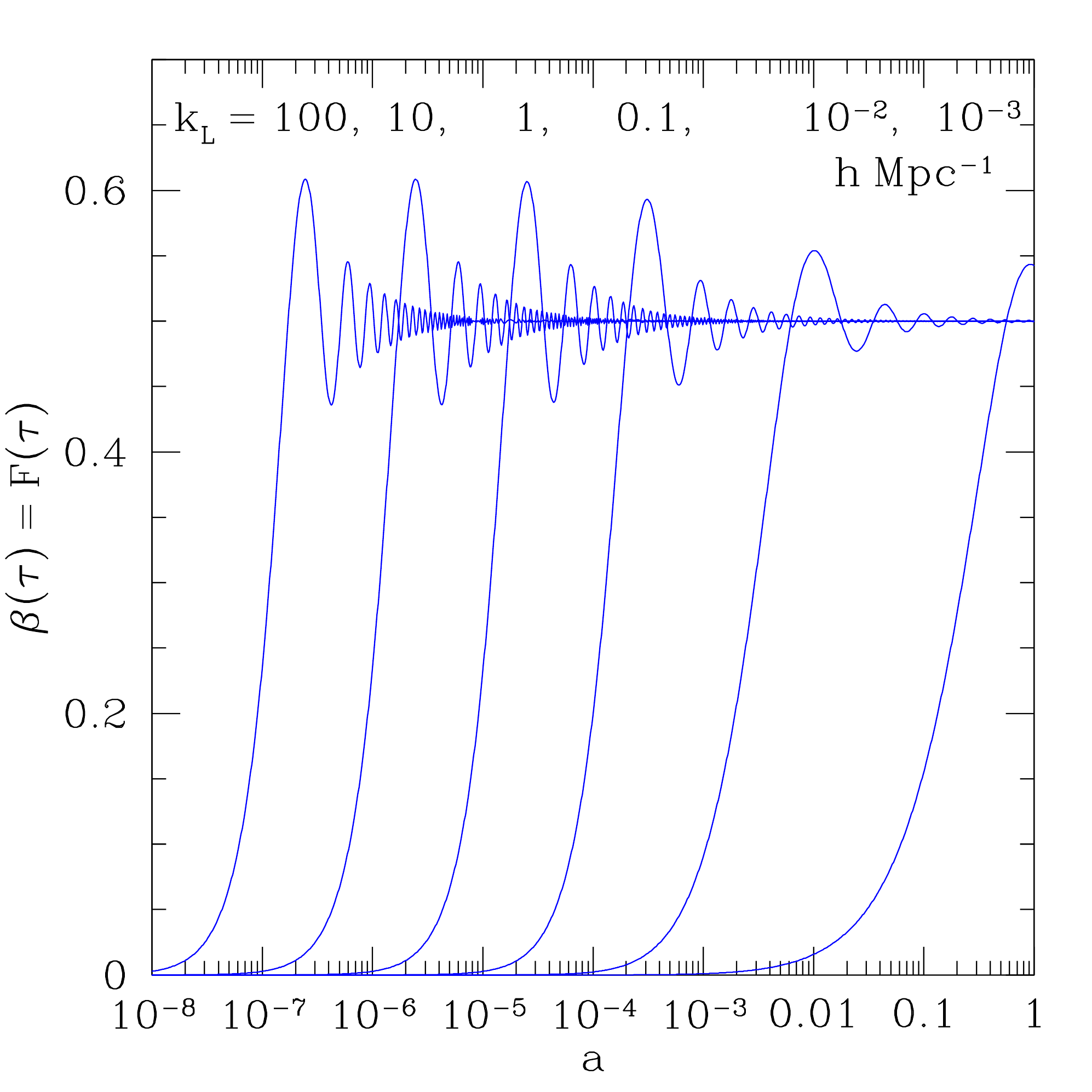}
\includegraphics[width=0.48\textwidth]{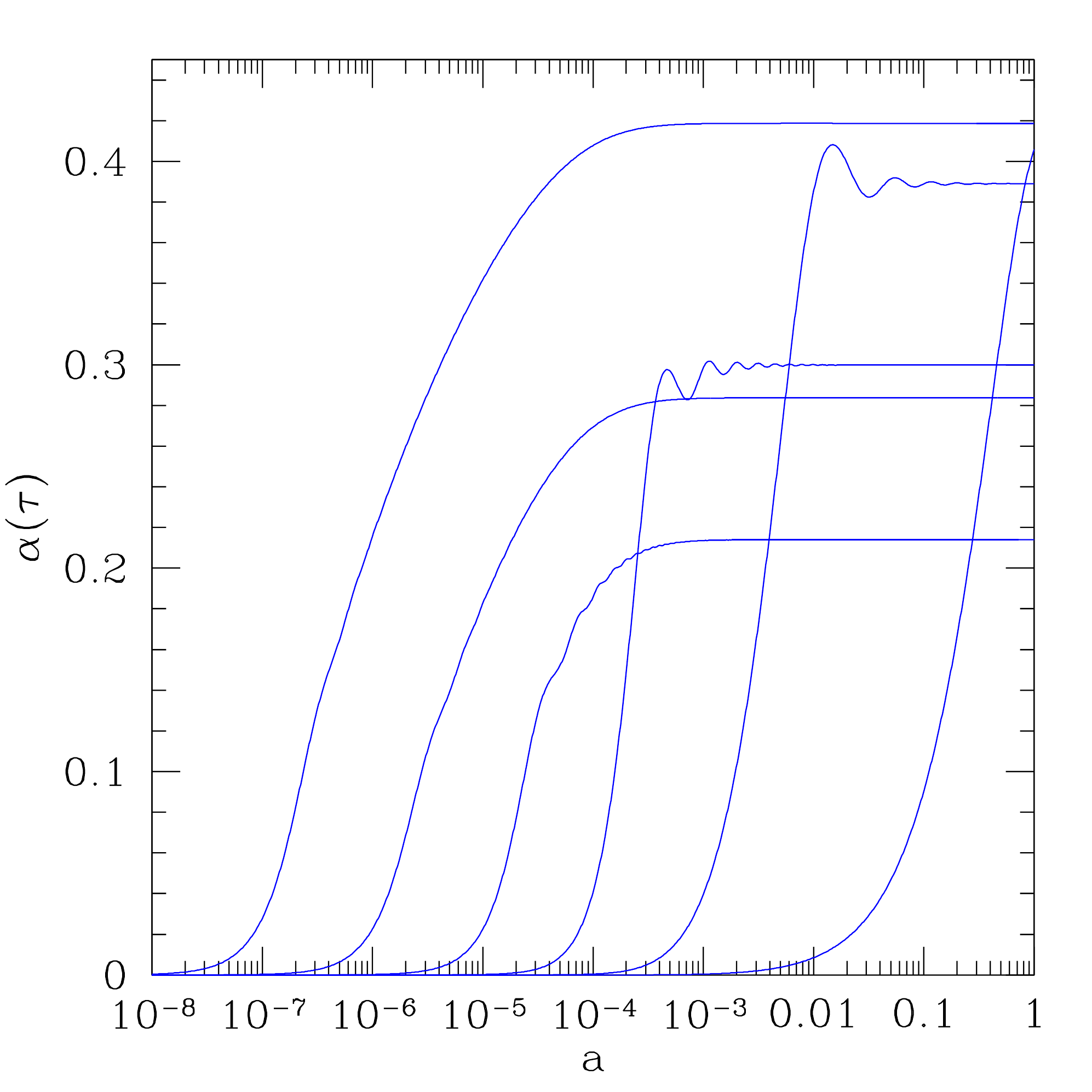}
\caption{\textit{Top panel:} the function $F(\tau)$ [\refeq{FVtensor}],
corresponding to $\beta(\tau)$ in \refeq{d2ttensor}, for tensor
modes of various wavenumbers $k_L$ as function of scale factor.  
\textit{Bottom panel:} coefficient $\alpha(\tau)$ in \refeq{d2ttensor},
as a function of $a$ for tensor modes with the same wavenumbers $k_L$ as
in the upper panel.  For wavenumbers entering the horizon during matter 
domination, this reduces to \refeq{alphaMD}.  
All results were obtained by numerical integration of the linear
and second order equations (see \refapp{numerics}) for a flat 
$\Lambda$CDM cosmology.  
}
\label{fig:FDsigma}
\end{figure}
%!!!!!!!!!!!!!!!!!!!!!!!!!!!!!!!!!!!!!!!!!!!
%%%%%%%%%%%%%%%%%%%%%%%%%%%%%%%%%%%%%%%%%%%%%%%%%%%%%%%%%%%%%%%%%%%%%%%%%%%
\subsection{Second order density}
\label{sec:tensord}

\refeq{tij} yields a tidal transfer function of
\be
T(\tau) = -\frac12 a^{-1} (a D_h')' \,,
\ee
so that 
\ba
F(\tau) =\:& - \frac12 \left[D_h(\tau)-1\right]
\vs
V(\tau) =\:& - \frac12 \int_0^\tau d\tau' \frac{\tau'^5}{\tau^5} D_h'(\tau')\,,
\label{eq:FVtensor}
\ea
where we have used that $D_h(0) = 1$.  
Inserting this into \refeq{d2tLPT}, and making use of
$h_i^{\  i} = 0$ yields the second order density induced by tensor or
vector mode tidal fields at time $\tau$, in terms of the linear small-scale 
density field
$\d_{1,s}$ and the \emph{primordial} tensor mode amplitude, or
vector mode at production, respectively, $h_{ij}^{(0)}$:
\ba
\d_{2,t}(\vx,\tau) =\:& h_{ij}^{(0)}(\vx) \left[ \alpha(\tau) \frac{\partial^i\partial^j}{\nabla^2} + \beta(\tau) x^i \partial^j \right] \d_{1,s}(\vx,\tau)
\label{eq:d2ttensor}
\ea
\ba
\alpha(\tau) =\:& \frac45 \left\{ F(\tau) - V(\tau) \right\}
\label{eq:alphaMD}\\
=\:& \frac{2}{5}+18 \frac{\cos(k_{L}\tau)}{(k_{L}\tau)^{4}}+6\frac{\sin(k_{L}\tau)}{(k_{L}\tau)^{3}}\left[  1-\frac{3}{(k_{L}\tau)^{2}}\right]\,,\nonumber\\
\beta(\tau) =\:& F(\tau)\label{eq:beta}\\
=\:&\frac{1}{2}+3\frac{\cos(k_{L}\tau)}{2(k_{L}\tau)^{2}} +3 \frac{\sin(k_{L}\tau)}{2(k_{L}\tau)^{3}}\,.\nonumber
\ea
This is \refeq{d2tintro} and the main analytical result of the paper.  In
order to elucidate its properties, we will consider tensor modes of fixed
wavenumber $k_L$ (we will briefly discuss vector modes below).  
In the limit of $k_L\tau \to 0$, that is before the
tensor mode enters the horizon, we have
\be
F(\tau) \to \frac1{20} (k_L\tau)^2; \quad
V(\tau) \to \frac1{70} (k_L\tau)^2\,,
\ee
so that \refeq{d2ttensor} becomes
\ba
\d_{2,t}(\vx,\tau) \stackrel{k_L\tau\to 0}{=}\:& (k_L\tau)^2
h_{ij}^{(0)} \frac15\left[ \frac1{7} \frac{\partial^i\partial^j}{\nabla^2}  
+ \frac14 x^i \partial^j
\right] \d_{1,s}(\vx, \tau)\,.
\ea
Thus, the lowest order effect of a long-wavelength tensor mode with
wavenumber $k_L$ comes in at order $(k_L\tau)^2$, as expected.

A further interesting limit to consider is $k_L\tau \gg 1$, that is long after
horizon crossing of the tensor mode.  In this limit, $D_h \to 0$, signifying
that the tensor mode has decayed away.  Thus, 
$V(\tau) \to 0$ while $F(\tau) \to 1/2$.  \refeq{d2ttensor} becomes
in this ``fossil'' limit
\ba
\d_{2,t}(\vx,\tau) \stackrel{k_L\tau \to \infty}{=}\:& 
h_{ij}^{(0)}\left[ \frac25 \frac{\partial^i\partial^j}{\nabla^2}  
+ \frac12 x^i \partial^j \right] \d_{1,s}(\vx, \tau)\,.
\label{eq:d2tfossil}
\ea
The first term in \refeq{d2tfossil} is the tidal interaction in the strict 
sense of the word. The second term corresponds to the effect of
an anisotropic expansion rescaling the physical coordinates. This term has the same form as that derived by \cite{MasuiPen}, the difference being that here this is just one specific limit of a more general $k_{L}$-dependent expression.  
One way of visualizing this term is to imagine a set of freely falling test particles in a 
Universe that is unperturbed apart from the tensor mode, which are initially
(at $k_L\tau \ll 1$) arranged as a spherical shell $\vx^2 = r^2$.   Then, 
their distribution at a later time is given by
\be
\vx^2 - 2 F(\tau) h_{ij}^{(0)}x^i x^j = r^2\,,
\ee
that is the particles have been rearranged into an ellipsoid.  
Note that the coefficient $F(\tau)$ of this term is valid for a general expansion 
history, provided the corresponding transfer function $D_h(\tau)$ is used in \refeq{Fdef}.  
On the other hand, the tidal interaction term in \refeq{d2ttensor} does depend on the background
cosmology; for example, $V(\tau)$ depends on the behavior of the decaying
scalar mode.  

Finally, we point out that \refeq{d2ttensor} agrees with the results
of App.~D in \cite{dai/etal:13}, which were derived for matter domination
as well.  Specifically, from Eq.~(D19) in that paper we
see that their function $2 \mathcal{S}_N(K)$ is equal to $\alpha$ as 
defined in \refeq{alphaMD}.

The functions $\beta(\tau) = F(\tau)$ and $\alpha(\tau)$ are shown 
in the upper and lower panel, respectively, of \reffig{FDsigma}
as function of scale factor for various tensor mode wavenumbers $k_L$.  
As expected, $F(\tau)$ asymptotes to $1/2$ long after horizon entry
of the tensor mode, while $\alpha(\tau)$ asymptotes to $2/5$ for modes
entering the horizon during matter domination as assumed in this derivation
(see curve for $k_L = 0.01 \iMpch$ in the lower panel of \reffig{FDsigma}).  
We will consider the case of tensor modes entering during radiation ($k_L \gg 0.01 \iMpch$) and 
$\Lambda$ domination ($k_L \ll 0.01 \iMpch$) in the next section.

\begin{figure}[t!]
\centering
\includegraphics[width=0.48\textwidth]{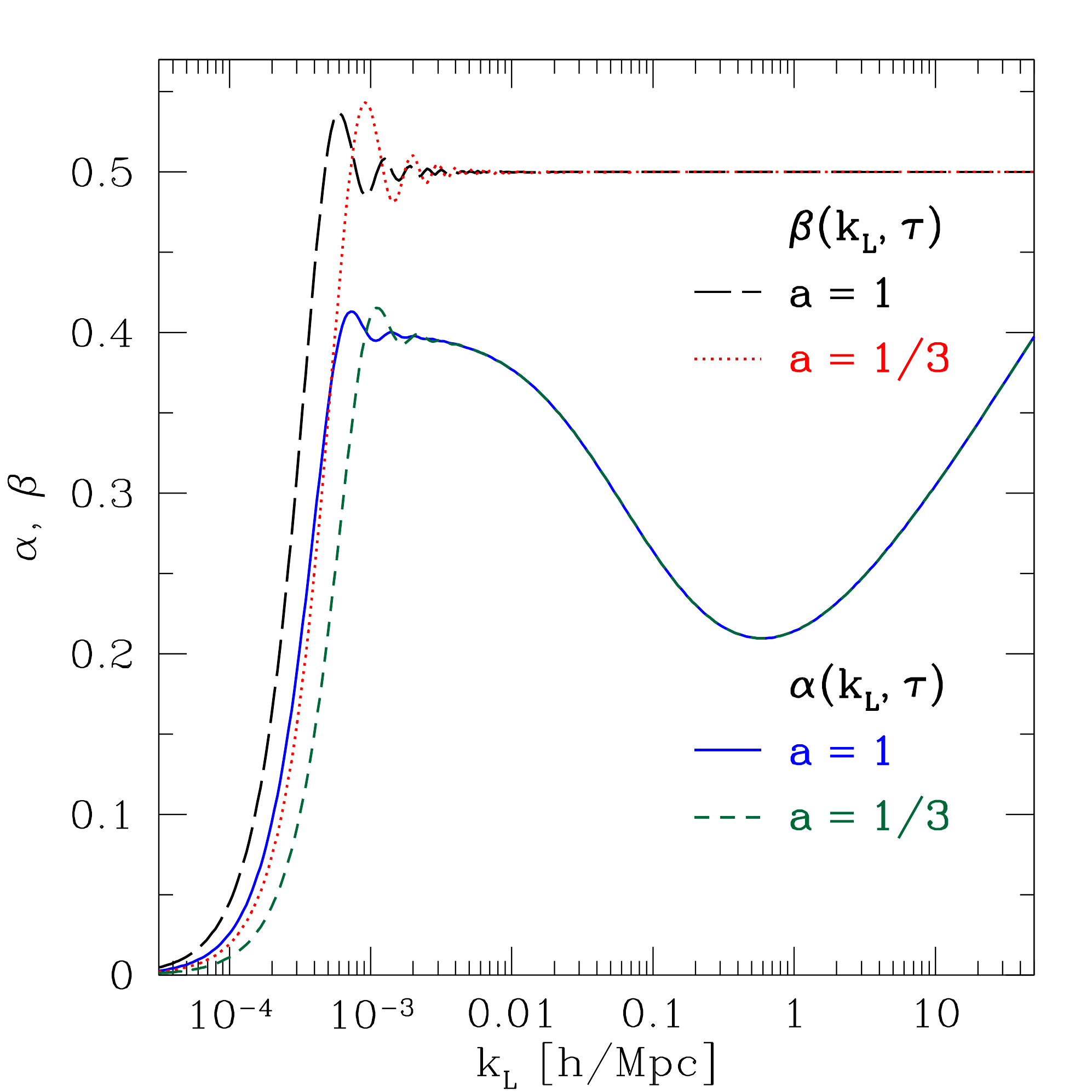}
\caption{Coefficient functions $\alpha(k_L,\tau),\,\beta(k_L,\tau)$ in \refeq{d2ttensor} as function of $k_L$
for fixed scale factors $a(\tau) = 1$ and $a(\tau) = 1/3$.  The same
$\Lambda$CDM cosmology as in \reffig{FDsigma} was assumed.}
\label{fig:d2t}
\end{figure}

\reffig{d2t} shows $\alpha(k_L,\tau),\,\beta(k_L,\tau)$ vs $k_L$ for two different values of the
scale factor.  Modes of wavenumber approaching $\cH$ 
have not decayed yet resulting in the oscillatory features 
around $k_L \sim 10^{-3} h\,{\rm Mpc}^{-1}$.  The effect of modes of even 
longer wavelength ($k_L \ll 10^{-3} h\,{\rm Mpc}^{-1}$) is strongly suppressed, 
since they have not entered the current horizon yet.  On the other hand, modes
with $k_L \gg \cH$ asymptote to a redshift-independent value, which for 
$\beta$ is $1/2$ at all wavenumbers, while for $\alpha$ this value is $2/5$
for a narrow range of scales for which the approximation of matter domination
applies.

%!!!!!!!!!!!!!!!!!!!!!!!!!!!!!!!!!!!!!!!!!!!
%!!!!!!!!!!!!!!!!!!!!!!!!!!!!!!!!!!!!!!!!!!!

%%%%%%%%%%%%%%%%%%%%%%%%%%%%%%%%%%%%%%%%%%%%%%%%%%%%%%%%%%%%%%%%%%%%%%%%%%%
\subsection{Including radiation and $\Lambda$}
\label{sec:tensorRD}

The analytical results derived in the previous sections apply only
for tensor modes which enter the horizon during matter domination.  
In order to extend the applicable range in $k_L$, we perform a numerical
integration of the linear and second order equations, including radiation
and a cosmological constant.  The details are given in \refapp{numerics}.  
\reffigs{FDsigma}{d2t} in fact show these numerical results.  Note that
we have completely neglected the effects of the baryon-photon fluid 
before recombination.  Thus, the results are only strictly valid for
scalar perturbations with wavelength shorter than the dissipation scale for which the 
perturbations in the baryon-photon fluid are erased by viscosity.

%!!!!!!!!!!!!!!!!!!!!!!!!!!!!!!!!!!!!!!!!!!!
\begin{figure}[t!]
\centering
\includegraphics[width=0.48\textwidth]{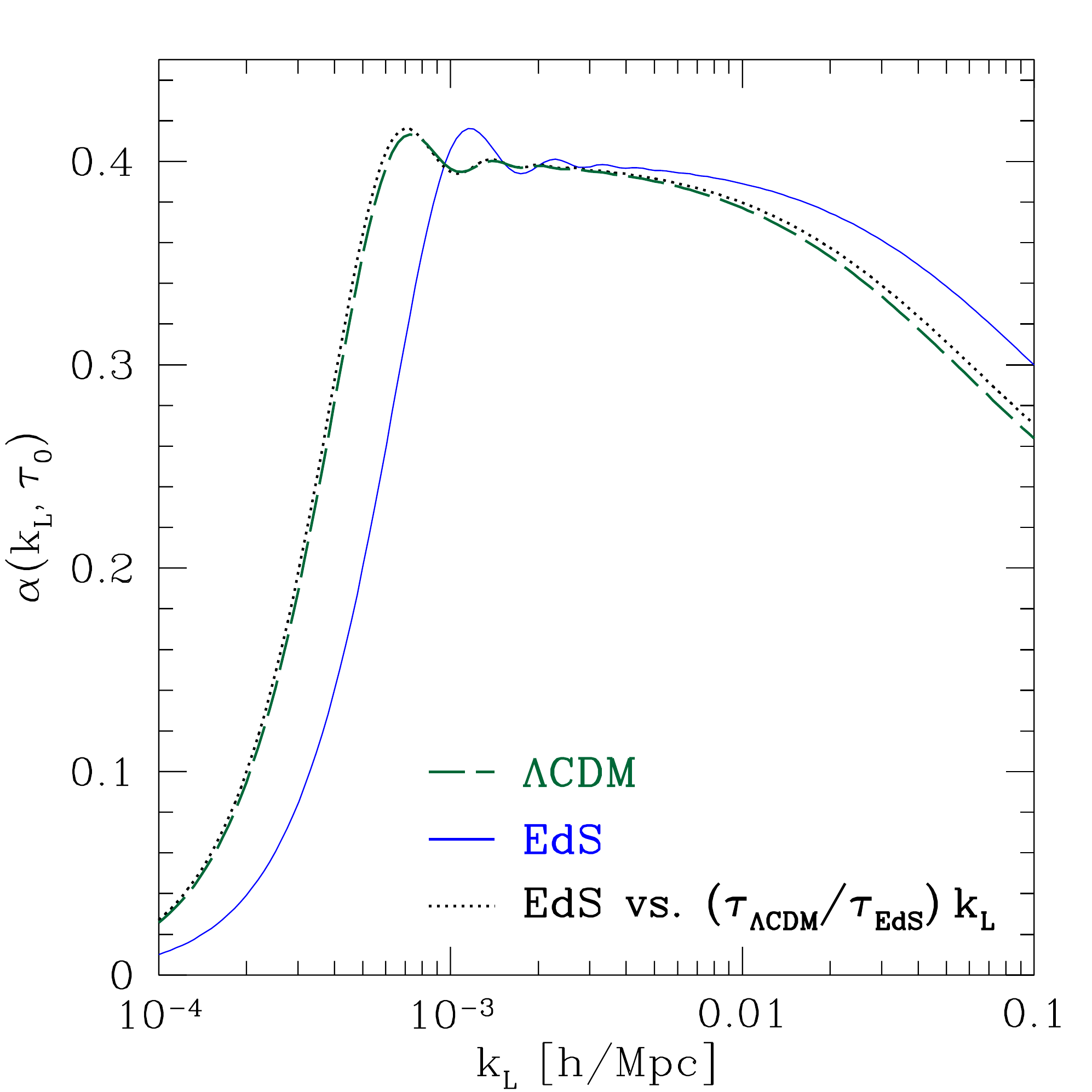}
\caption{Coefficient function $\alpha(k_L,\tau_0)$ as in \reffig{d2t} 
(green dashed), and for a flat matter dominated (Einstein-de Sitter, EdS) 
Universe (blue solid).  The black dotted line shows the EdS result
plotted at the same $k_L\tau$ as $\Lambda$CDM (see text).  All curves are for
$a(\tau_0)=1$.}
\label{fig:d2tLambda}
\end{figure}
%!!!!!!!!!!!!!!!!!!!!!!!!!!!!!!!!!!!!!!!!!!!

Let us first consider the effects of radiation, important for tensor modes
with $k_L \gtrsim 0.01 h {\rm Mpc}^{-1}$.  On intermediate scales, there is a suppression
of the second order density field, while on very small scales, we see a 
logarithmic increase of the effect.  It is possible to qualitatively 
understand these trends through an analytical derivation in pure
radiation domination, analogous to \refsec{tensord}, which is described in
\refapp{RD}.  
During radiation domination, both scalar and tensor modes evolve differently
than during matter domination, with the scalar growth being proportional
to $\ln \tau \propto \ln a$ rather than $a(\tau)$.  The derivation of
\refsec{LPT} can easily be adapted to this case (again 
assuming that the scalar perturbation is in the growing mode), leading to 
the result
\ba
\alpha_{\rm RD}(\tau) =\:& V^{\rm RD}(\tau) \vs
\beta_{\rm RD}(\tau) =\:& F(\tau)\,,
\ea
where the coefficient function $V^{\rm RD}(\tau)$ is defined in \refeq{VRDdef}
while $F(\tau)$ is still given by \refeq{FVtensor} (with the appropriate
 $D_h$ for radiation domination).  
Compared to the result for matter domination, note that the contribution 
to $\alpha(\tau)$, $4/5\, F(\tau)$ from the second order
density (fourth term in \refeq{dtLPT1}) has canceled with a contribution of
opposite sign in the contribution from second order evolution.  
This leads to the suppression on intermediate scales relative to the
matter domination result.  The second difference to the case of matter 
domination is that  $V^{\rm RD}(\tau)$ continues to evolve logarithmically 
even as $k_L\tau \gg 1$, explaining the trend seen for 
$k_L \geq 10 h {\rm Mpc}^{-1}$ in \reffig{d2t}.  

We now turn to the effects of $\Lambda$.  The main effect of
$\Lambda$ is to modify the linear scalar growth factor and the scale 
factor-conformal time relation.  This
is illustrated in \reffig{d2tLambda}, where we also show the result for
a Universe with $\Lambda=0$, with rescaled $k_L$ (dotted line) 
so as to show results
at the same value of $k_L\tau$ in both cases.  We see agreement to better than
5\%.  Note that in the $\Lambda=0$ case we have rescaled radiation to
keep $\Omega_{r 0}/\Omn$ and thus $a_{\rm eq}$ fixed.

In summary, while the presence of radiation and cosmological constant 
change the results in detail, the key result is that $\alpha$ remains
in the range of $0.2-0.5$ over the entire range of accessible 
tensor wavenumbers, and the range of redshifts relevant for large-scale
structure.  Thus, we do not confirm the expectation of
\cite{dai/etal:13} who argued that the effect should be strongly suppressed
for modes $k_L \gtrsim 0.01 \iMpch$.

%%%%%%%%%%%%%%%%%%%%%%%%%%%%%%%%%%%%%%%%%%%%%%%%%%%%%%%%%%%%%%%%%%%%%%%%%%%
%%%%%%%%%%%%%%%%%%%%%%%%%%%%%%%%%%%%%%%%%%%%%%%%%%%%%%%%%%%%%%%%%%%%%%%%%%%
\section{Impact on small-scale density statistics}
\label{sec:xilocal}

Consider a region of size $R \ll 1/k_L$ over which a tensor mode $h_{ij}$ 
with wavenumber $k_L$ can
be considered spatially constant (as described in \refsec{FNC}, 
this region is entirely contained within the conformal Fermi patch 
corresponding to the tensor mode).  Corrections will be suppressed by
powers of $k_L/k_S$, where $k_S$ is the wavenumber of the scalar perturbation
considered.  We now show how the results for the second-order density 
$\d_{2,t}$ from the previous section predict a modification of the small-scale 
correlation function $\xi_F(\vr)$ of the matter density field
measured locally, that is in the \fncb~frame.

We define the local two-point function in \fncb~$\xi_F(\vr,\tau)$ through
\be
\xi_F(\vr,\tau) = \left\< \d_F(\v{0},\tau) \d_F(\vr,\tau) \right\>\,,
\label{eq:xihdef}
\ee
where
\be
\d_F(\vx,\tau) = \d_{1,s}(\vx,\tau) + \d_{2,t}(\vx,\tau)\,.
\ee
Analogously, we define the linear matter correlation function $\xi_{1,s}$, 
i.e. that of $\d_{1,s}$.  
By assumption $r \ll 1/k_L$.  Inserting \refeq{d2ttensor} yields
\ba
\xi_F(\vr,\tau) =\:& \xi_{1,s}(r,\tau) \vs
& + h_{ij}^{(0)}\left[ 2 \alpha(\tau) \frac{\partial^i\partial^j}{\nabla^2} 
+ \beta(\tau) r^i \partial_j \right] \xi_{1,s}(r,\tau) \vs
%%%
= \xi_{1,s}(r,\tau) +\: &
 h_{ij}^{(0)} \hat r^i \hat r^j\bigg[ 2 \alpha(\tau)
\left(\xi_{\nabla^{-2}\d_{1,s}}''(r) - \frac1r \xi_{\nabla^{-2}\d_{1,s}}'(r) \right) \vs
& \hspace*{1.2cm} + \beta(\tau) \frac{d \xi_{1,s}(r,\tau)}{d\ln r} \bigg] \,,
\label{eq:xitensor}
\ea
where primes denote derivatives with respect to $r$ and
\be
\xi_{\nabla^{-2} \d_{1,s}}(r,\tau) = -\int \frac{d^3\vk}{(2\pi)^3} k^{-2} P_{1,s}(k,\tau) e^{i \vk\vr}\,.
\ee
Note that the same result is obtained when defining \refeq{xihdef} as $\<\d_F(-\vr/2)\d_F(\vr/2)\>$.  
We can Fourier transform \refeq{xitensor} to obtain the local power spectrum.  
This parallels the calculation in App.~B of \cite{conformalfermi} and yields
\ba
P_\d(\vk,\tau|h) - P_\d(k,\tau) =\:& \hat k^i \hat k^j h_{ij}^{(0)} P_\d(k,\tau)\vs
&\times \left[ 2 \alpha(\tau) - \beta(\tau) \frac{d\ln P_\d(k,\tau)}{d\ln k} \right] \,.
\nonumber
\ea
This is \refeq{Pkintro}.

\refeq{xitensor} applies to the matter density field.  In reality, one 
observes the statistics of some tracer of matter which is generally biased,
i.e. whose clustering properties are not identical to those of matter.  
The simplest case is a linear local bias.  Then, \refeq{xitensor} remains 
valid for the tracers if $\xi_{1,s}$ is replaced with
\be
\xi_{1,s,g}(r) = b_1^2\:\xi_{1,s}(r)\,,
\ee
where $b_1$ is the linear bias.  
Two conditions need to be fulfilled for this to be valid:  first, the matter 
correlation $\xi_{1,s}(r)$ on the scale $r$ has to be small, so that
corrections of order $[b_2 \xi_{1,s}(r)]^2$ are negligible, where $b_2$
is the quadratic bias of the tracer.  Second, any non-locality in the 
relation between tracer number density and matter density has to be 
restricted to scales much smaller than $r$.  These two conditions are 
likely to be satisfied by the 21cm emission from the dark ages which
was considered in \cite{MasuiPen,jeong/kamionkowski:2012}.  On the other
hand, for galaxy surveys at lower redshifts, these conditions are not met 
in general, necessitating a higher order perturbative treatment, which we
will not attempt here.  

%%%%%%%%%%%%%%%%%%%%%%%%%%%%%%%%%%%%%%%%%%%%%%%%%%%%%%%%%%%%%%%%%%%%%%%%%%%
\subsection{Projection effects}
\label{sec:proj}

\refeq{xitensor} gives the effect of tensor modes
on small-scale correlations that would be observed locally by a comoving
observer, i.e.~in the observer's \fncb~frame.  In the end however, we 
want to know the effect on the small-scale correlations as observed
from Earth, specifically through the arrival directions and redshifts
of photons.  We thus need to add the effects of the
mapping from \fncb(source location) to \fncb(Earth), which we refer to
as ``projection effects''.\footnote{As long as the region
over which we measure small-scale correlations is much smaller than the
horizon, which is always the case in practice, the distinction between
\fnc~and \fncb~is irrelevant (\refsec{FNC}).}  
Note that since the two frames are uniquely
defined (at the relevant order), these projection effects are gauge-invariant.  
As described in Sec.~IV of \cite{conformalfermi}, there are three ingredients
to this mapping:
\begin{itemize}
\item The transformation of the correlation scale $\vr$ through the
cosmic ruler perturbations derived in \cite{stdruler}.
\item A shift in time from fixed proper time at \fncb(source) to fixed observed redshift on Earth, as derived in \cite{Tpaper}.
\item A modulation of the mean observed density within the patch over which the
small-scale correlation is measured.  The modulation of the observed density
of a general tracer by tensor modes was derived in \cite{GWpaper}.
\end{itemize}

As shown in \refapp{proj}, it is straightforward to collect all these
results to obtain
\ba
 \xi(\vr, \tau|h) &= \Bigg[  1 + \left\{\frac12 \int_0^{\chit} d\chi\: h_{\parallel}' \: \d_{ij} + \frac12 h_{ij} + \partial_{(i} \D x_{j)}\right\}  r^i \partial_{r}^j \vs
 +\:& \frac1{2\cH} \int_0^{\chit} d\chi\: h_{\parallel}' \:\partial_\tau  
\label{eq:xiproj}\\
 +\:& 2 \left\{ \frac{b_e}2\int_0^{\chit} d\chi\: h_{\parallel}' + \partial_i \D x^i 
+ \Q \:\M_T \right\}\Bigg] \xi_F(\vr;\tau)\,,
\nonumber
\ea
where again $\xi_F(\vr,\tau)$ is the small-scale correlation function in the
\fncb~frame.  
Here, metric perturbations outside integrals are evaluated at the source, while
metric perturbations inside integrals are evaluated on the past lightcone in 
the background (see \refapp{proj} for details).  Further, 
$h_\parallel \equiv h_{ij}\nhat^i \nhat^j$, $\vnhat$ is the unit vector along the
line of sight, and
$\chit \equiv \chib(\zt)$ where $\chib(z)$ is the comoving distance-redshift 
relation in the background and $\zt$ is the observed redshift.  
$\D x^i$, defined in \refeqs{Dxpar}{Dxperp} denote the displacement of the
true source position from the apparent position.  $\M_T$ is the gauge-invariant
magnification produced by the tensor mode and is given in \refeq{MT}.  
Further, there are two tracer-dependent parameters: the 
magnification bias parameter $\Q$, given in the simplest case of a sharp
flux-limited survey by $\Q = - d\ln \bar{n}_g/d\ln f_{\rm cut}$;  and the
paramater $b_e$, which quantifies the redshift evolution of the comoving
number density of tracers through \refeq{btdef}.

We now insert \refeq{xitensor} for $\xi_F$ in \refeq{xiproj}, which effectively
results in adding up all tensor mode effects.  We then obtain
\ba
 \xi(\vr, \tau|h) =& \Bigg[  1 + \left\{\frac12 \int_0^{\chit} d\chi\: h_{\parallel}' \: \d_{ij} + \frac12 h_{ij} + \partial_{(i} \D x_{j)}\right\}  r^i \partial_{r}^j \vs
& + \frac1{2\cH} \int_0^{\chit} d\chi\: h_{\parallel}' \:\partial_\tau  
\label{eq:xitot}\\
& + 2 \left\{ \frac{b_e}2\int_0^{\chit} d\chi\: h_{\parallel}' + \partial_i \D x^i 
+ \Q \:\M_T \right\} \vs
&+ 2 h_{ij}^{(0)}\left\{ \alpha(\tau) \frac{\partial^i\partial^j}{\nabla^2} 
+ \beta(\tau) r^i \partial_j \right\} \Bigg] \xi_{1,s}(r;\tau)\,.
\nonumber
\ea
For biased tracers, we again just need to replace $\xi_{1,s}$ with
$b_1^2 \xi_{1,s}$ as long as linear local bias is sufficient.  This
result agrees with \cite{dai/etal:13}, with the exception of the magnification
bias contribution $\Q \M_T$ which they did not include.

%%%%%%%%%%%%%%%%%%%%%%%%%%%%%%%%%%%%%%%%%%%%%%%%%%%%%%%%%%%%%%%%%%%%%%%%%%%
%%%%%%%%%%%%%%%%%%%%%%%%%%%%%%%%%%%%%%%%%%%%%%%%%%%%%%%%%%%%%%%%%%%%%%%%%%%
\section{Shear correlation and intrinsic alignments}
\label{sec:IA}

We now turn to the signature of tensor modes in galaxy shear surveys.  
As discussed in \cite{DodelsonEtal,GWshear}, the effect of tensor modes
on photon geodesics leads to a contribution to the weak lensing shear
which can be measured through the correlations of galaxy shapes (second
moments).  Specifically, a gravitational wave background contributes to
the parity-odd B-mode component, which does not receive any scalar contribution
at linear order and thus provides a window to search for gravitational
waves.  

The tensor-mode contribution to the shear, which is a tracefree symmetric
tensor on the sky (with two independent components) is given by \cite{GWshear}
\be
\g^{\rm proj}_{ij} = -\frac12 \P_i^{\  k}\P_j^{\  l} h_{kl} - \partial_{\perp (i} \D x_{\perp j)}\,,
\label{eq:shear}
\ee
where $\P_{kl} = \d_{kl} - \nhat_k \nhat_l$ is the projection operator onto the sky plane, 
$\partial_{\perp i} = \P_i^{\  j}\partial_j$, and the displacement $\D x_\perp$
is given in \refeq{Dxperp}.  

\refeq{shear} gives the contribution to galaxy shape correlations induced
by the mapping from the source's \fncb~frame to observed coordinates, in
analogy with the projection effects discussed in \refsec{proj} [in fact,
$\g_{ij}$ is part of the terms appearing in the first line of \refeq{xiproj}].  
However, we expect that the
effect of tensor modes on the local density field described by 
$\d_{2,t}$ [\refeq{d2ttensor}] also affects galaxy shapes;  that is, there
is also a correlation of galaxy shapes with the local tidal field in the 
\fncb~frame, leading to a contribution to shape correlations which we will 
denote as $\g^{\rm IA}_{ij}$.  In the terminology of weak lensing shear, this
effect is referred to as \emph{intrinsic alignment}.  Just as in \refsec{proj},
the \emph{observed} correlation of galaxy shapes is then given by the
sum of the two contributions,
\be
\g^{\rm obs}_{ij} = \g^{\rm proj}_{ij} + \g^{\rm IA}_{ij}\,.
\ee  
We now consider the intrinsic alignment contribution in more detail.  It
is difficult to predict the amplitude of the alignment of galaxy shapes with
the large-scale tidal field from first principles.  For tensor modes,
this effect was first considered by \citet{GWshear}, who used observations
of alignments with scalar tidal fields to estimate the effect for gravitational
waves.  Using our results from \refsec{tensor}, we can elaborate a bit more
on this effect.  We have found above that the second order density field induced
by the (trace-free component of) a \emph{scalar} tidal field is given by
\be
\d_{2,s} = \frac27 \left(\frac{\partial_i \partial_j}{\nabla^2} \d_{1,L}(\vx,\tau) \right) \frac{\partial^i \partial^j}{\nabla^2} \d_{1,s}(\vx, \tau)\,,
\ee
while that of a tensor mode is given by
\be
\d_{2,t} = \alpha(k_{L}, \tau) h_{ij}^{(0)} \frac{\partial^i \partial^j}{\nabla^2} \d_{1,s}(\vx, \tau)\,.
\ee
One possible way to estimate the intrinsic alignment by tensor modes is
to assume that the alignment scales as the second order density perturbation
induced by the external tidal field.  The alignment by scalar tidal
fields has observationally been measured at low redshift for elliptical
galaxies \cite{HirataEtal:07,OkumuraJing,BlazekEtal}.  The scalar 
intrinsic  alignment contribution to the shear can be parametrized as
\be
\gamma^{\rm IA,s}_{ij}(\vx,\tau) = -\Omn \tilde C a^{-1}(\tau_P) \P_i^{\  k} \P_j^{\  l} \left(\frac{\partial_k \partial_l}{\nabla^2} \d(\vx,\tau_P) \right)\,,
\nonumber
\ee
where $\tau$ is the observation epoch while $\tau_P$ is the epoch at which
the tidal field is evaluated.  Note that since scalar tidal fields do not
evolve strongly, the precise value of $\tau_P$ does not change results by more
than 30\%.  In the following, we will choose $\tau_P = \tau$.  
For the $z=0.2-0.4$ luminous red galaxies (LRG) studied in \cite{OkumuraJing,BlazekEtal}, $\tilde C = C_1\rho_{\rm cr,0} \approx 0.12$.  Thus, by matching the 
second order density induced by scalar and tensor tidal fields at the 
observation epoch, we arrive at the following estimate for the tensor 
contribution to intrinsic alignments:
\be
\gamma^{\rm IA,t}_{ij}(\vx,\tau) = - \Omn \tilde C\: \frac72 \frac{\alpha(k_{L}, \tau)}{a(\tau)}
\P_i^{\  k} \P_j^{\  l} h_{kl}^{(0)}(\vx) \,.
\label{eq:gammaIAt}
\ee

%!!!!!!!!!!!!!!!!!!!!!!!!!!!!!!!!!!!!!!!!!!!
\begin{figure}[t!]
\centering
\includegraphics[width=0.49\textwidth]{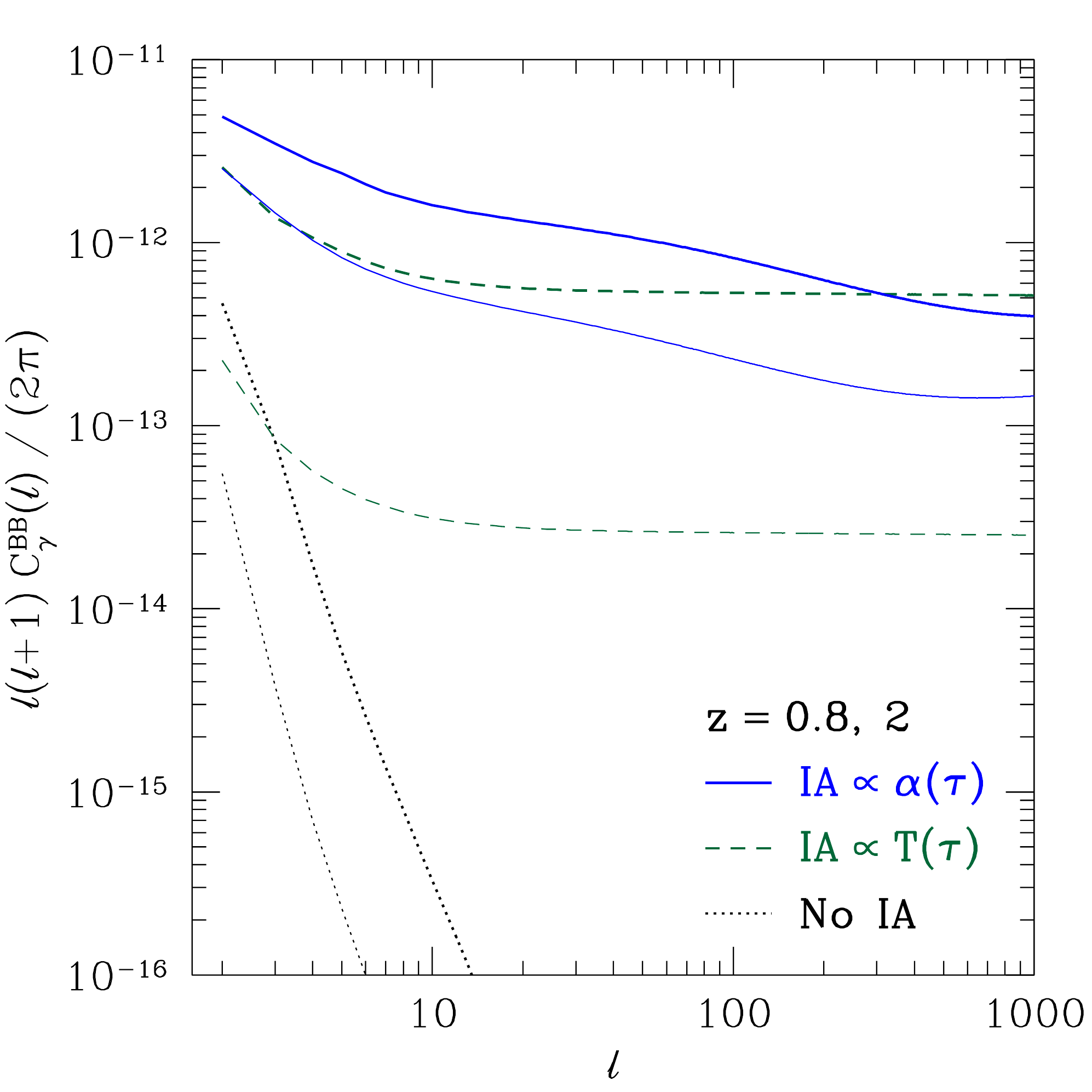}
\caption{Tensor mode contribution to the $B$ mode angular shear 
power spectrum from a gravitational wave background with tensor-to-scalar 
ratio $r=0.1$.  The blue solid lines show the result using
the matching of the second order density [\refeq{gammaIAt}],  
green dashed lines show the result when using the instantaneous tensor tidal
field as adopted in \cite{GWshear}, and black dotted
lines show the result in the absence of intrinsic alignment, i.e. only
including the lensing contribution.  In all cases, thick lines are for
a source redshift of $z=2$, while thin lines are for $z=0.8$.}
\label{fig:Cl}
\end{figure}
%!!!!!!!!!!!!!!!!!!!!!!!!!!!!!!!!!!!!!!!!!!!

\reffig{Cl} shows the resulting predicted angular power spectrum of the
B-mode of the shear, assuming an almost scale-invariant gravitational wave
background with tensor-to-scalar ratio $r=0.1$ (that is, exactly one half
of the value adopted in \cite{GWshear}, with otherwise identical parameters).  
The solid lines lines show the result obtained using the matching relation \refeq{gammaIAt} at observed redshift $z = 0.8$ (thin) and $z=2$ (thick), respectively.  
Here, we have used the numerical results for $\Lambda$CDM (\refapp{numerics}) for 
the coefficient $\alpha(k_L, \tau)$, as shown in \reffig{d2t}.
The dotted lines in \reffig{Cl} show the result for $\tilde C=0$, i.e. when only
the lensing (projection) effect contributes and any alignment effect of 
the tensor mode tidal field is absent.  Clearly, the lensing effect is at
least one order of magnitude smaller than the estimated alignment effect,
and moreover drops much more rapidly towards higher $\ell$ (smaller scales).  

The long-dashed lines in \reffig{Cl} show the previous prescription adopted in \cite{GWshear},
which relates the shear to the instantaneous tidal field induced by tensor
modes,
\be
\gamma^{\rm IA,t}_{ij}(\vx,\tau) = \tilde C\: \P_i^{\  k} \P_j^{\  l} t_{kl}^{\rm tensor}(\vx, \tau)\,,
\label{eq:gammaIAprev}
\ee
for the same redshifts.  
Unlike \refeq{gammaIAt}, this relation only yields a contribution to the shear
for tensor modes that have recently entered the horizon.  For a relatively
high source redshift of $z=2$, the prediction from the second order density-matching is a factor of $2-4$ 
larger than the prediction using \refeq{gammaIAprev} on large scales, well within the
uncertainty of these rough estimates, while at lower source 
redshifts the difference becomes much larger (factor of $\sim 30$ at $z=0.8$).  
This is because the effect has more time to build up in \refeq{gammaIAt},
i.e. a wider range in wavenumber contributes to the alignment signal.  
Consequently, in this ansatz the redshift-dependence of the alignment 
contribution is significantly weaker (it is almost entirely due to the
$a^{-1}(\tau)$ prefactor in \refeq{gammaIAt}, given the redshift independence 
of the tidal imprint on the second order density field for $k_L \gg \cH$).  
The redshift dependence is weaker at the lowest $\ell$, because very large 
scale tensor modes that have recently entered the horizon have had less time 
to build up an effect at earlier times.  Finally, we point out that the
redshift dependence of the alignment strength of galaxies is very uncertain
at this point and likely depends strongly on the particular galaxy sample
considered.

Nevertheless, the fact that the signal is much larger at low source redshifts than what
was estimated in \cite{GWshear} greatly improves the observational
prospects for detecting this effect.  
Note also the slow suppression of the signal for $l \gtrsim 100$ of the 
alignment signal predicted by \refeq{gammaIAt}, which is due to the suppression
of the second order density in the range $0.01 < k_L [h/{\rm Mpc}] < 1$
[\reffig{d2t}];  however, given the order-of-magnitude uncertainty of this estimate
such details of the predicted signal should be taken with a grain of salt.

Finally, in our prediction of $\g^{\rm IA}_{ij}$, we have only considered
the first term in \refeq{d2tscalar} and \refeq{d2ttensor}, respectively.  
There might also be a contribution from the differential displacement
[second term in \refeq{d2ttensor} and last two terms in \refeq{d2tscalar}]
to the orientation of galaxies in both scalar and tensor cases, 
but investigating the relative contribution of this term
goes beyond the scope of this paper.

%%%%%%%%%%%%%%%%%%%%%%%%%%%%%%%%%%%%%%%%%%%%%%%%%%%%%%%%%%%%%%%%%%%%%%%%%%%
%%%%%%%%%%%%%%%%%%%%%%%%%%%%%%%%%%%%%%%%%%%%%%%%%%%%%%%%%%%%%%%%%%%%%%%%%%%
\section{Conclusions}
\label{sec:concl}

In this paper we have computed the effect of long-wavelength perturbations on the dynamics of short-wavelength matter inhomogeneities. We made crucial use of conformal Fermi Normal Coordinates (\fncb) in order to isolate the physical effects and remove unobservable coordinate artifacts. This ensures that the results of the various steps of our computations are individually observable. Our formalism can be applied to scalar, vector and tensor long-wavelength perturbations. The case of scalars is well-known and provides a nice check of our results. For vector and tensor perturbations we find that the effect on the short-wavelength matter inhomogeneity is given by \refeq{d2tintro}, with the consequent anisotropy in the power spectrum given by \refeq{Pkintro}.  Interestingly, these effects remain of order $h_{ij}^{(0)}$ in the $k_L\tau \gg 1$ (fossil) limit, even for wavenumbers $k_L$ which have entered the horizon during radiation domination.  

All these results are given in the reference frame of a comoving observer. We have also computed projection effects necessary to make contact with what is observed from Earth. The projected 2-point function of matter (or linearly biased tracers) is given in \refeq{xitot} as an example, but other projected quantities can be straightforwardly computed as well.  
Although we have not done so, this can also be easily coupled to the
signal-to-noise forecasts for 21cm correlations performed in \cite{BookMKFS,jeong/kamionkowski:2012}.  Further, it would be interesting to study other
large-scale structure tracers as probes of this effect, and cross-correlations between
them and the cosmic microwave background.

A further important pertinent observable is cosmic shear, i.e.~correlations of galaxy shapes. Here, the locally observable (``tidal'') effects and projection effects are commonly known as ``intrinsic alignments'' and ``lensing'', respectively.  We have made the rough approximation that the tidal alignment of galaxies scales as the anisotropic contribution to the second order density field.  The resulting odd parity B-mode power spectrum, to which there are no scalar contributions at linear order, is shown in \reffig{Cl}.  Interestingly, the tidal effects are much stronger than the lensing effects for tensor modes, while the converse is true for standard scalar density perturbations.  Moreover, we have found that the residual effect in the density field which remains when $k_T\tau \gg 1$ [\refeq{d2tfossil}] greatly enhances the expected signal for source galaxies at low redshifts ($z < 2$).  

The assumption that galaxy alignments scale with the anisotropic part of
the second order density field can only be seen as a rough approximation.  
It would thus be interesting to perform N-body simulations with an external
tidal field imposed, which mimics the effect of a long-wavelength
gravitational wave.  
One could then study the alignment of dark matter halos and substructure
with this external tidal field.

Another approximation we have made throughout is to neglect the 
effects of the baryon-photon fluid before recombination.  This issue can
be studied using the same tools as presented here.  However, since the
fluid is relativistic, one also has to take into account the metric
components $g_{0i}^F$ and $g_{ij}^F$.  We leave this for future work as well.

%%%%%%%%%%%%%%%%%%%%%%%%%%%%%%%%%%%%%%%%%%%%%

\section*{Acknowledgments} 
We are happy to thank L.~Dai, D.~Jeong and M.~Kamionkowski for helpful discussions.  We further thank K.~Masui for discussions and pointing out a factor of 2 mistake in \refsec{xilocal}.  E.~P.~was supported in part by the Department of Energy grant DE-FG02-91ER-40671.  F.~S.~acknowledges support at Princeton by NASA through Einstein Postdoctoral Fellowship grant number PF2-130100 awarded by the Chandra X-ray Center, which is operated by the Smithsonian Astrophysical Observatory for NASA under contract NAS8-03060.  M.~Z.~is supported in part by the National Science Foundation grants PHY-0855425, AST-0907969, PHY-1213563 and by the David $\&$ Lucile Packard Foundation.

%%%%%%%%%%%%%%%%%%%%%%%%%%%%%%%%%%%%%%%%%%%%%%%%%%%%%%%%%%%%%%%%%%%%%%%%%%%%%%

\begin{widetext}

\appendix

%%%%%%%%%%%%%%%%%%%%%%%%%%%%%%%%%%%%%%%%%%%%%%%%%%%%%%%%%%%%%%%%%%%%%%%%%%%%%
%%%%%%%%%%%%%%%%%%%%%%%%%%%%%%%%%%%%%%%%%%%%%%%%%%%%%%%%%%%%%%%%%%%%%%%%%%%%%
\section{Derivation of \refeqs{FNCmetric}{tij}}
\label{app:FNC}

This section deals with the transformation of the metric from a set of global coordinates \refeq{metrichij} to \fncb, specifically the time-time component.  We will set $h_{0i}=0$ and only keep $h_{00}$ and $h_{ij}$, since this is sufficient to treat scalar perturbations in all widespread
gauge choices as well as tensor perturbations.  The relation between global
coordinates and \fncb~is then \cite{conformalfermi}
\ba
x^0(\bar x^\alpha_F) =\:& \bar x^0_F + \frac12 \int_0^{\bar x^0_F} h_{00}(\tau) d\tau 
+ u_i  \bar x_F^i  - \frac14  h'_{ij} \bar x_F^i \bar x_F^j + \O[(\bar x_F^i)^3] 
\label{eq:FNCbtrans0}
\\
%%%
x^k(\bar x^\alpha_F) =\:& u^k (\bar x_F^0 - \tau_F) + \bar x_F^k - \frac12 h^k_{\  i} \bar x_F^i  - \frac14 \left[h^k_{\  i,j} + h^k_{\  j,i} - h_{ij}^{\  \  ,k}
\right] \bar x_F^i \bar x_F^j + \O[(\bar x_F^i)^3]\,,
\label{eq:FNCbtransi}
\ea
where $u^i$ is the coordinate velocity of the central geodesic, and all perturbations 
are evaluated at the central geodesic, i.e. at $\bar\vx_F = 0$.  
First, let us verify that using this coordinate transform, the conformal
metric $\eta_{\mu\nu} + h_{\mu\nu}$ becomes $\eta_{\mu\nu} + \O([\bar x_F^i]^2)$, 
i.e. Eq.~(14) in \cite{conformalfermi}.  We have
\ba
\frac{\partial x^0}{\partial\bar x_F^0} =\:& 1 + \frac12 h_{00}(\v{0},\bar x_F^0) + u_i' \bar x_F^i \vs
\frac{\partial x^i}{\partial \bar x_F^0} =\:& u^i\,.
\label{eq:FNCderiv}
\ea
Neglecting terms of order $(u/c)^2$, the central point
has a four-velocity
\be
\uu^\mu = \frac{dP^\mu}{dx^0} = \left(1 + \frac12 h_{00},\:u^i \right)
\ee
which follows the geodesic equation for the metric $\eta_{\mu\nu}+h_{\mu\nu}$ whose spatial components are
\be
u'^i = - \Gamma^i_{00} (1 + \O(h))^2 = \frac12 \partial^i h_{00}\,.
\ee
We are interested in the time-time component of the conformal metric, which we denote $\bar g_{00}$.  In global coordinates, it is given by
\be
\bar g_{00}(x) = \eta_{00} + h_{00}(x) = -1 + h_{00}(\v{0},\tau) + \partial_i h_{00}(\v{0},\tau) x^i
+ \frac12 \partial_i \partial_j h_{00}(\v{0},\tau) x^i x^j + \O[(x^i)^3]\,,
\ee
where we have chosen the global coordinate origin to coincide with that
of the \fncb~frame at the specific time considered.  
In going to \fncb, the second term is immediately canceled by the
second term in \refeq{FNCbtrans0}, while the third, gradient term is
canceled by the $u^i$ term in \refeq{FNCbtrans0} together with the 
geodesic equation.  Thus, we obtain for the 00 component of the conformal
metric in \fncb~coordinates
\be
\bar g_{00}^F = \eta_{00} + \frac12 \partial_i \partial_j h_{00}\,\bar x_F^i \bar x_F^j + \frac12 h_{ij}'' \bar x_F^i \bar x_F^j\,,
\ee
where again $h$ is evaluated at $\v{0}$.  
Here, we have replaced $x^i \to \bar x_F^i$ at this order.  This
is Eq.~(14) of \cite{conformalfermi}.  

We now consider physical metric, that is $g_{\mu\nu} = a^2(\tau)[\eta_{\mu\nu} + h_{\mu\nu}]$.  
First, the origin of the \fncb~frame is now constrained
to follow a geodesic in the physical metric.  The four-velocity is given
by (e.g., \cite{stdruler})
\be
\uu^\mu = \frac{dx^\mu}{d\lambda} = a^{-1}(x^0) \left(1 + \frac12 h_{00}(x),\: u^i(x) \right)\,.
\ee
The relevant Christoffel components are
\ba
\Gamma^i_{00} =\:& \frac12 a^{-2} \left[- a^2 h_{00}^{\  \   ,i} \right]
= -\frac12  \partial^i h_{00}
\vs
\Gamma^i_{0j} =\:& \frac12 a^{-2} \left[ g^i_{\  j,0} \right]
= \frac{a'}{a} \d^i_{\  j} + \O(h) = \cH \d^i_{\  j} + \O(h)\,.
\ea
Since $\Gamma^i_{0j}$ multiplies $u$, we do not have to include the $h$ terms
there.  We then obtain for the $i$-component of $\uu^\mu$
\ba
\frac{d}{d\lambda} \left(\frac{ u^i}{a}\right) 
= (\uu^0) \left(\frac{ u^i}{a}\right)'
=\:& - \Gamma^i_{00} (\uu^0)^2 - 2\Gamma^i_{0j} \uu^0 \uu^j 
= a^{-2} \left[\frac12 \partial^i h_{00} - 2 \cH  u^i\right]
\vs
\Rightarrow \; u'^i =\:& \frac12 \partial^i h_{00} - \cH  u^i\,.
\label{eq:geod}
\ea
Note that this agrees with the Euler equation in \refeq{euler} once linearized.  
\refeq{FNCderiv} is still valid, since $u^i$ in \refeq{FNCderiv} is
defined as $dP^i/dx^0$ and 
\be
\frac{dP^i}{d x^0} = \frac{dx^i/d\lambda}{dx^0/d\lambda} = \frac{\uu^i}{\uu^0} =  u^i\,,
\ee
recalling that $u^i$ is first order.   We now transform $g_{00}$ to $g_{00}^F$ 
in \fncb.  Using \refeq{geod}, we obtain
\be
g_{00}^F = a^2\left(x^0[\bar x_F]\right) \left[-1 + 2 \cH u_i \bar x_F^i + \frac12 \partial_i \partial_j h_{00}\,\bar x_F^i \bar x_F^j + \frac12 h_{ij}'' \bar x_F^i \bar x_F^j\right]\,.
\ee
We expand the prefactor $a^2\left(x^0[\bar x_F]\right)$ as follows:
\ba
a^2\left(x^0[\bar x_F]\right) = a^2\left(\bar x_F^0 + \frac12\int h_{00} d\tau\right)
\left[1 + 2 \cH u_i \bar x_F^i - \frac12 \cH h_{ij}' \bar x_F^i \bar x_F^j
\right]\,.
\ea
Putting everything together, the $00$ component of the physical metric
in \fncb~becomes
\ba
g_{00}^F =\:& a_F^2(\bar x_F^0)
\left[-1 + \frac12 \partial_i \partial_j h_{00}\,\bar x_F^i \bar x_F^j + \frac12 h_{ij}'' \bar x_F^i \bar x_F^j + \frac12 \cH h_{ij}' \bar x_F^i \bar x_F^j
\right]
\label{eq:g00F} \\
a_F(\bar x_F^0) \equiv\:& a\left(\bar x_F^0 + \frac12\int h_{00}(\v{0},\tau) d\tau\right)\,.
\label{eq:aF}
\ea
This corresponds to \refeqs{FNCmetric}{tij}.  
$g_{00}^F$ is, apart from the scale factor, clearly in the FNC form, with
corrections going as spatial distance from the central geodesic squared.  
The fact that the scale factor is evaluated at $\bar x_F^0 + \frac12\int h_{00}(\v{0},\tau) d\tau$ might surprise at first.  Note however that by construction,
at $\bar\vx_F=0$ the coordinate $\bar x_0^F$ is the proper time along the
central geodesic with respect to the metric $\eta_{\mu\nu} + h_{\mu\nu}$, so that
\be
t_F \equiv \int^{\bar x_F^0} a\left[x^0(\tau, \bar\vx_F=0)\right] d\tau
= \int^{\bar x_F^0} a_F(\tau) d\tau 
\ee
is the proper time (with respect to the physical metric $g_{\mu\nu}$) along 
the central geodesic.  Thus, evaluating \refeq{g00F} at $\bar\vx_F=0$ yields
\be
dt_F = a_F(\bar x_F^0) d\bar x_F^0\,.
\ee
In other words, $\bar x_F^0$ has a clear interpretation as 
``conformal proper time''.  In particular for $h_{00}=0$, $t_F = t$, where $t$
is the time coordinate, and $\bar x_F^0 = \tau$, where $\tau$ is the
corresponding conformal time.  For $h_{00} \neq 0$, the non-trivial
argument of the scale factor in \refeq{aF} expresses the fact that
constant-proper-time surfaces are \emph{not} constant-scale-factor
surfaces.  If we transform to the standard FNC frame, i.e. to physical
rather than comoving coordinates, we obtain
metric corrections of the form $H^2 \vx_F^2$, $\dot H \vx_F^2$, where
$H$ is the Hubble rate evaluated at the time coordinate corresponding to
the given proper time along the central geodesic, i.e. at 
$\bar x_F^0 + \frac12 \int h_{00}(\v{0},\tau)d\tau$ just as in the
scale factor above.  The apparent unphysical dependence on a metric
perturbation $h_{00}$ (without any derivative) is simply because we are referring
to an unobservable ``background'' scale factor here.  A local observer
moving along the central geodesic will simply measure the Hubble
rate as a function of his/her proper time (this is in fact is how we 
define our background scale factor in practice).  Thus, the scale factor 
multiplying the metric \refeq{g00F} is the scale factor that 
would \emph{locally be reconstructed} from the measured Hubble rate,
hence our notation of $a_F(\bar x_F^0)$ in \refeq{FNCmetric}.

%%%%%%%%%%%%%%%%%%%%%%%%%%%%%%%%%%%%%%%%%%%%%%%%%%%%%%%%%%%%%%%%%%%%%%%%%%%
%%%%%%%%%%%%%%%%%%%%%%%%%%%%%%%%%%%%%%%%%%%%%%%%%%%%%%%%%%%%%%%%%%%%%%%%%%%
\section{Eulerian derivation}
\label{app:eulerian}

In this section we present an independent derivation of the main results
of \refsec{LPT} using Eulerian perturbation theory.  We define the peculiar 
velocity through
\be
\v{u} = \frac{d\vx}{d\tau} = a(t) \frac{d\vx}{dt} = a \v{v} - \cH \vx,\quad \v{v} = \frac{d\vr}{dt}\,.
\label{eq:udef}
\ee
The continuity and Euler equations for an ideal fluid are then given by
\ba
\d' + \vn\cdot\left[(1+\d) \v{u}\right] =\:& 0 
\label{eq:cont}\\
%%%
\v{u}' + (\v{u}\cdot\vn) \v{u} + \cH \v{u} =\:& -\vn \Phi - \frac1\rho \vn\left( \rho \v{\s}\right)\,,
\label{eq:euler}
\ea
where $\s_{ij}$ is the stress tensor of the fluid including pressure (here, $[\vn(\rho\v{\s})]^i = \partial_j(\rho \sigma^{ij})$).  In the following, we will set $\s_{ij}=0$.  
Separating density and velocity into parts zeroth and first order in 
$t_{ij}$, we write
\ba
\d = \d_s+\d_t; \quad \v{u} =\:& \v{u}_s + \v{u}_t\,,
\ea
where $\d_s,\,\v{u}_s; \d_t,\,\v{u}_t$ satisfy
\ba
\d_s' + \vn\cdot[(1+\d_s)\v{u}_s] =\:& 0 \vs
\v{u}'_s + (\v{u}_s\cdot\vn) \v{u}_s + \cH \v{u}_s =\:& -\vn \Phi_s
\label{eq:eulers} \\
%%%
\d_t' + \vn\cdot\v{u}_t + \vn\cdot[\d_s\v{u}_t + \d_t \v{u}_s] =\:& 0 \vs
\left[\v{u}'_t + (\v{u}_s\cdot\vn) \v{u}_t + (\v{u}_t\cdot\vn) \v{u}_s + \cH \v{u}_t\right]^i =\:& - T(\tau) t^{(0)\,i}_{\quad\    j} x^j \,.
\label{eq:eulerH}
\ea
The linearized equations are easily seen to be equivalent to the corresponding
Lagrangian equations.  Hence, we can make use of \refeqs{sLs}{dLt} for the
linear solutions.  In particular, 
\ba
u_{1,s}^i(\vx,\tau) =\:& - a'(\tau) \frac{\partial^i}{\nabla^2} \d_{1,s}(\vx,\tau_0) \vs
u_{1,t}^i(\vx,\tau) =\:& -F'(\tau) t_{ij}^{(0)} x^j \vs
\d_{1,t}(\vx,\tau) =\:& F(\tau) t_{i}^{(0)\,i}\,.
\ea

% % % % % % % % % % % % % % % % % % % % % % % % % % % % % % % % % % % % % % %
\subsection{Second order solution}

As in the Lagrangian derivation (\refsec{LPT}), we work in a perturbative expansion in all of $\d_s,\,\v{u}_s; \d_t,\,\v{u}_t$, i.e.
\ba
\v{u}_t =\:& \v{u}_{1,t} + \v{u}_{2,t} + \cdots\,,
\ea
and derive the leading corrections $\d_{2,t},\,\v{u}_{2,t}$.  
These corrections obey the equations
\ba
\d_{2,t}' + \vn\cdot\v{u}_{2,t} =\:& - \vn\cdot[\d_{1,s}\v{u}_{1,t} + \d_{1,t} \v{u}_{1,s}]
\label{eq:contH1}\\
\v{u}_{2,t}' + (\v{u}_{1,s}\cdot\vn) \v{u}_{1,t} + (\v{u}_{1,t}\cdot\vn) \v{u}_{1,s} + \cH \v{u}_{2,t} =\:& -\vn\Phi_s^{(2)}\,.
\label{eq:eulerH1}
\ea
Taking the divergence of the second equation, assuming Einstein-de Sitter, 
and introducing 
$\theta_{2,t} = \vn\cdot \v{u}_{2,t}$ allows us to write these equations as
\ba
\d_{2,t}' + \theta_{2,t} =\: - S_1 \vs
\theta_{2,t}' + \cH \theta_{2,t} + \frac32 \cH^2 \d_{2,t} =\:& - S_2 \vs
S_1 =\:& -\left[ a F' t^{(0)}_{ij} x^i \partial^j 
+ (a F)' t^{(0)\,i}_i \right] \d_{1,s}^{(0)}\vs
S_2 =\:& \vn\cdot\left[ (\v{u}_{1,s}\cdot\vn) \v{u}_{1,t} + (\v{u}_{1,t}\cdot\vn) \v{u}_{1,s} \right] \vs
=\:& a' F' t_{ij}^{(0)} \left[ 2\frac{\partial^i\partial^j}{\nabla^2} \d_{1,s}^{(0)} + x^i \partial^j \d_{1,s}^{(0)} \right]\,.
\ea
Here, $\d_{1,s}^{(0)}$ stands for the linear scalar density at a reference time
$\tau_0$, i.e. with the growth taken out.  
We now take the derivative with respect to $\tau$ of the continuity
equation and insert the Euler equation for $\theta_{2,t}'$.  This yields
\ba
\d_{2,t}'' + \cH \d_{2,t}' - \frac32 \cH^2 \d_{2,t} = - [S_1' + \cH S_1] + S_{2,t} = -\frac1a [a S_1]' + S_2\,,
\label{eq:d2eomcomp3}
\ea
which can also be written as
\ba
\d_{2,t}'' + \frac2\tau \d_{2,t}' - \frac6{\tau^2} \d_{2,t} =\:&  -\frac1a [a S_1]' + S_2
\vs
=\:& t_{ij}^{(0)} \left\{
\left[ 3 a' F' + a F'' \right] x^i \partial^j 
+ 2 a' F' \frac{\partial^i \partial^j}{\nabla^2} 
+ \frac1a \left[a (a F)' \right]' \d^{ij}
\right\}
\d_{1,s}^{(0)}
\,.
\ea
The Green's function for this ODE, with the boundary conditions $\d_{2,t}(0) = \d_{2,t}'(0) = 0$, is
\be
G(\tau,\tau') = \frac15 \left(\frac{\tau^2}{\tau'} - \frac{\tau'^4}{\tau^3}\right) \Theta(\tau-\tau')\,.
\ee
We then have
\ba
\int_0^\tau d\tau' G(\tau,\tau') a' F' =\:& \frac25 a(\tau) [F(\tau)-V(\tau)] \vs
\int_0^\tau d\tau' G(\tau,\tau') a F'' =\:& \frac15 a(\tau) [-F(\tau) + 6 V(\tau)] \vs
\int_0^\tau d\tau' G(\tau,\tau') \frac1a \left[a (a F)'\right]' =\:& 
a(\tau) \left[ F(\tau) + \frac32 D_{\s2}(\tau) \right]\,.
\ea
Putting everything together, we obtain
\ba
\d_{2,t}(\vx,\tau) =\:& t_{ij}^{(0)} \left\{
F(\tau) x^i \partial^j 
+ \frac45 \left[F(\tau)-V(\tau)\right] \frac{\partial^i \partial^j}{\nabla^2} 
+ \left[F(\tau) + \frac32 D_{\s2}(\tau) \right] \d^{ij}
\right\}
\d_{1,s}(\vx,\tau)\,.
\ea
This agrees exactly with the result of the Lagrangian
derivation, \refeq{d2tLPT}.

%%%%%%%%%%%%%%%%%%%%%%%%%%%%%%%%%%%%%%%%%%%%%%%%%%%%%%%%%%%%%%%%%%%%%%%%%%%%%%
%%%%%%%%%%%%%%%%%%%%%%%%%%%%%%%%%%%%%%%%%%%%%%%%%%%%%%%%%%%%%%%%%%%%%%%%%%%%%%
\section{Numerical evaluation for $\Lambda$CDM}
\label{app:numerics}

This section presents the equations used to numerically
evaluate the tensor mode contribution for $\Lambda$CDM, i.e. including
a cosmological constant in addition to matter and radiation.  
The numerical results are shown
in \reffigs{FDsigma}{d2tLambda}.  We have (assuming flatness)
\ba
H^2 = H_0^2 \left[ \Omega_{\Lambda 0} + \Omn a^{-3} + \Omega_{r 0} a^{-4} \right] =: H_0^2 E^2(a)\,,
\label{eq:HLCDM}
\ea
where $a=1$ today and 
\be
\Omega_{\Lambda 0} + \Omn + \Omega_{r 0} = 1\,.
\ee
The $\Omega_{X 0}$ refer to fractions of the critical density today.  
The relation between $a$ and the dimensionless conformal time 
$y = \tilde H_0 \tau$ needs to be solved numerically through
\be
y(a) = \int_0^a \frac{da'}{a'^2 E(a')}\,.
\ee
However, deep in radiation and matter domination, we use the
analytical result obtained when neglecting the $\Lambda$ term
in \refeq{HLCDM}:
\ba
y(a) =\:& 2\Omn^{-1/2} \left(\sqrt{a+\aeq} - \sqrt{\aeq}\right)
\,.
\ea

% % % % % % % % % % % % % % % % % % % % % % % % % % % % % % % % % % % % % %
\subsubsection{Linear solutions} 

Transforming the tensor mode equation to $y$, we obtain
\ba
D_h''(y) + 2 f_H(y) D_h'(y) + \frac{k_L^2}{\tilde H_0^2} D_h(y) = 0 
\label{eq:Dheom}\\
f_H(y) = a(y) E(a(y))\vs
D_h(0) = 1;\quad D_h'(0) = 0\,,
\ea
where for the remainder of this section primes stand for derivatives
with respect to $y$.  The Poisson equation is now
\be
\nabla^2 \Phi = \frac32 \Omn H_0^2 a^{-1}(y) \d(y)
\,.
\ee
We define a linear growth factor $D_{1,s}$ for scalar perturbations through
\be
\s_{1,s}(\vq,\tau) = D_{1,s}(\tau) \s_{1,s}(\vq,\tau_0)\,,
\ee
which satisfies
\ba
D_{1,s}''(y) + f_H(y) D_{1,s}'(y) - \frac32 \Omn a^{-1}(y) D_{1,s}(y) = 0
\ea
with boundary conditions
\be
D_{1,s}(0) = 0; \quad D_{1,s}(y_0) = 1\,.
\ee
In order to enforce these boundary conditions, we integrate the growth
equation from some $y_{\rm min}$ deep in radiation domination, with
initial conditions
\be
D_{1,s}(y_{\rm min}) = C; \quad D_{1,s}'(y_{\rm min}) = \frac{C}{y_{\rm min}}\,,
\ee
and adjust $C$ so that $D_{1,s}(y_0) = 1$ where $a(y_0)=1$.  

% % % % % % % % % % % % % % % % % % % % % % % % % % % % % % % % % % % % % %
\subsubsection{Second order solution} 

In order to solve for $\s_{2,t}$, we transform \refeq{s2t1} from $\tau$
to $y$, yielding
\ba
\s_{2,t}''(\vq,y) + f_H(y) \s_{2,t}'(\vq,y) 
- \frac32 \Omn a^{-1}(y)  \s_{2,t}(\vq,y) =\:& \Sigma(\vq,y) \vs
\Sigma(\vq,y) =\:& -\frac12 a^{-1} \frac{d}{dy} \left[ a \frac{d D_h}{dy}\right] D_{1,s}(y)   \Sigma_0(\vq) \vs
\Sigma_0(\vq) =\:& \left(\frac{\partial^i\partial^j}{\nabla^2} \d_{1,s}(\vq,y_0)\right) h_{ij}^{(0)}
\,.
\label{eq:s2eom}
\ea
As before, we start integrating at $y_{\rm min}$ where $k_L \tau_{\rm min} = k_L/\tilde H_0 \:y_{\rm min}$ is sufficiently small so that the right-hand side 
can be set to zero.  The initial conditions for $\s_{2,t}$ are then
\be
\s_{2,t}(y_{\rm min}) = 0;\quad \s_{2,t}'(y_{\rm min}) = 0\,.
\ee
We then solve \refeq{s2eom} numerically using a fourth-order Runge-Kutta
scheme with adaptive step size.

%%%%%%%%%%%%%%%%%%%%%%%%%%%%%%%%%%%%%%%%%%%%%%%%%%%%%%%%%%%%%%%%%%%%%%%%%%%%%%
%%%%%%%%%%%%%%%%%%%%%%%%%%%%%%%%%%%%%%%%%%%%%%%%%%%%%%%%%%%%%%%%%%%%%%%%%%%%%%
\section{Radiation domination}
\label{app:RD}

We now consider the case of dark matter during pure radiation domination (RD) 
in the Lagrangian treatment of \refsec{LPT}.  In addition to clarifying
the reason for the behavior of the tensor-scalar coupling shown in 
\reffig{d2t}, these results are also used for the initial conditions of
the numerical integration described in \refapp{numerics}.  We have
\be
\Om = 0; \quad
\H(a) = \tilde H_0 a^{-1}; \quad
\tau = \tilde H_0^{-1} a; \quad
\cH = \tau^{-1}\,,
\label{eq:RD}
\ee
where $\tilde H_0$ is the Hubble constant at some reference time during
RD where $a(\tau_0)=1$.  

% % % % % % % % % % % % % % % % % % % % % % % % % % % % % % % % % % % % % % %
\subsection{Linear solutions}

Since $\Om=0$, the $q$-divergence of the linearized scalar EOM becomes
\ba
\s_{1,s}''(\vq,\tau) + \frac1{\tau} \s_{1,s}'(\vq,\tau) =\:& 0\,.
\ea
The growing mode corresponds to $\s_{1,s} \propto \ln \tau$, while
the decaying mode is $\s_{1,s} = $~const.  In the following, we will
again assume that the scalar perturbations have settled in the growing
mode by the time when the tidal field $t_{ij}$ becomes relevant.  Since
the growth is only logarithmic in $\tau$ rather than polynomial as in
matter domination, this is a much stronger restriction.  

We normalize the density perturbation $\d_{1,s}(\vq,\tau)$ to its value
at horizon crossing $\cH_* = 1/\tau_* = k_S$:
\be
\d_{1,s}(\vq,\tau) = \ln (k_S \tau) \d_{1,s}^{\rm H}(\vq)\qquad (k_S\tau\gg1)\,,
\ee
which again is only valid if $k_S\tau \gg 1$.  We then have
\be
s_{1,s}^i(\vq,\tau) = - \ln (k_S \tau) 
\frac{\partial_q^i}{\nabla_q^2} \d_{1,s}^{\rm H}(\vq)\,.
\ee
For the tidal field, \refeqs{sLt}{dLt} are valid for a general expansion
history.

% % % % % % % % % % % % % % % % % % % % % % % % % % % % % % % % % % % % % %
\subsection{Second-order solution}

Again, since $\Om=0$, the source terms of the Poisson equation vanish, and 
\refeq{s2t1} simplifies to
\be
\s_{2,t}'' + \cH \s_{2,t}' = - \v{M}_{1,s}^{ij} t_{ij}\,,
\label{eq:s2t1RD}
\ee
where on the r.h.s. all contributions are evaluated at $\vq$ and $\tau$.  
In reality, $\Om$ is of course never exactly zero;  thus, our results assume
that $t^i_{\  i}$ is not dramatically enhanced so that
the prefactor of $\Om$ sufficiently suppresses the first source term in 
\refeq{s2t1} over the second one.  We obtain
\ba
\s_{2,t}''(\vq,\tau) + \frac1{\tau} \s_{2,t}'(\vq,\tau) =\:& \Sigma(\vq,\tau) \vs
\Sigma(\vq,\tau) = \ln (k_S \tau) T(\tau) \Sigma_0(\vq) =\:& 
\ln (k_S \tau) T(\tau)  \left(\frac{\partial^i\partial^j}{\nabla^2} \d_{1,s}^{\rm H}(\vq)\right) t_{ij}^{(0)}
\,.
\ea
The growing and decaying modes are again
$\s_{2,t} \propto \ln \tau$ and $\s_{2,t} \propto $~const, respectively.  
The solution for this equation with the appropriate boundary condtions 
is given by
\be
\s_{2,t}(\vq,\tau) = \int_0^\tau d\tau' \tau' \ln\left(\frac{\tau}{\tau'}\right) 
\Sigma(\vq,\tau')\,.
\ee
We then obtain using integration by parts (and $F(\tau\to0) = 0$) to obtain
\ba
\s_{2,t}(\vq,\tau) =\:& D_{\s1}(\tau) \Sigma_0(\vq) \\
%%%
D_{\s1}(\tau) =\:&  
\int_{1/k_S}^\tau d\tau' 
\tau' \ln\left(\frac{\tau}{\tau'}\right) \ln(k_S \tau') T(\tau') \vs
=\:& F(\tau) \ln(\tau k_S) - 2 \int_{1/k_S}^\tau \frac{d\tau'}{\tau'} F(\tau')\,.\nonumber
\ea
Thus,
\ba
\s_{2,t}(\vq,\tau) = t_{ij}^{(0)} \left(\frac{\partial^i\partial^j}{\nabla^2} \d_{1,s}(\vq,\tau)\right) \left[ F(\tau) - V^{\rm RD}(\tau) \right]\,,
\label{eq:s2tRD}
\ea
where $F$ is defined as before [\refeq{Fdef}] and we have introduced
\ba
V^{\rm RD}(\tau) =\:& 2\, [\ln k_S\tau]^{-1}\int_{1/k_S}^\tau \frac{d\tau'}{\tau'} F(\tau') \,.
\label{eq:VRDdef}
\ea
As in \refsec{LPT}, the total contribution to the Eulerian density induced by
the external tidal field is then given by \refeq{dtLPT1}, which yields for
the second order part
\ba
\d_{2,t}(\vx, \tau) =\:& t_{ij}^{(0)}\left[
 - \left[ F(\tau) - V^{\rm RD}(\tau)\right] \frac{\partial^i \partial^j}{\nabla^2}
+ F(\tau) \left(
\frac{\partial^i \partial^j}{\nabla^2} + \d^{ij} + x^i \partial^j \right)\right] 
\d_{1,s}(\vx, \tau) \vs
=\:& t_{ij}^{(0)}\left[
  V^{\rm RD}(\tau) \frac{\partial^i \partial^j}{\nabla^2}
+ F(\tau) \left( \d^{ij} + x^i \partial^j\right) \right] \d_{1,s}(\vx, \tau)\,.
\label{eq:d2tRD}
\ea
This has very similar structure to the result in matter domination
[\refeq{d2tLPT}], the key difference being that the coefficient of
the tidal term $\partial^i\partial^j/\nabla^2 \d_{1,s}$ only involves
the function $V^{\rm RD}$ rather than $F$.

%%%%%%%%%%%%%%%%%%%%%%%%%%%%%%%%%%%%%%%%%%%%%%%%%%%%%%%%%%%%%%%%%%%%%%%%%%%%%%
\subsection{Tensor modes}
\label{app:tensorRD}

As in \refsec{tensor}, we have $F(\tau) = \frac12 \left[1 - D_h(\tau)\right]$,
where in RD 
\be
D_h(\tau) = \frac{\sin k_L \tau}{k_L \tau}\,.
\ee
The function $V^{\rm RD}(\tau)$ becomes
\ba
V^{\rm RD}(\tau) =\:& [\ln k_S\tau]^{-1}\int_{k_L/k_S}^{k_L\tau} 
\frac{dx}x \left(1-\frac{\sin x}x\right) \vs
=\:& [\ln k_S\tau]^{-1} \left[ \ln x - {\rm Ci}\: x + \frac{\sin x}x \right]_{k_L/k_S}^{k_L \tau}\,.
\ea
In the $k_L\tau\to\infty$ limit, $F(\tau) \to 1/2$ just as in matter domination. On the other hand, $V^{\rm RD}(\tau)$ becomes
\ba
V^{\rm RD}(\tau) \to\:& [\ln k_S\tau]^{-1} 
\frac{\ln k_L\tau}{\ln k_S\tau} = \frac{\ln k_L\tau}{\ln k_L\tau + \ln(k_S/k_L)}\,.
\ea
For $k_L \tau \gg k_S/k_L$, $V^{\rm RD}$
logarithmically approaches 1 from below.  \refeq{d2tRD} becomes in this limit
\ba
\d_{2,t}(\vx,\tau) =\:& 
h_{ij}^{(0)}\left[ \frac{\ln k_L\tau}{\ln k_L\tau + \ln(k_S/k_L)} \frac{\partial^i\partial^j}{\nabla^2}  
+ \frac12 x^i \partial^j \right] \d_{1,s}(\vx, \tau)\,.
\label{eq:d2tfossilRD}
\ea
Thus, there is a non-zero second-order density for $k_L\tau\gg 1$
during radiation domination as well.  Moreover, the only difference
to the corresponding result for matter domination [\refeq{d2tfossil}]
is the numerical coefficient of the first term ($2/5$ in MD, 
order 1 in RD depending on $k_L\tau$), and the fact that it evolves
logarithmically with $\tau$.

%%%%%%%%%%%%%%%%%%%%%%%%%%%%%%%%%%%%%%%%%%%%%%%%%%%%%%%%%%%%%%%%%%%%%%%%%%%
%%%%%%%%%%%%%%%%%%%%%%%%%%%%%%%%%%%%%%%%%%%%%%%%%%%%%%%%%%%%%%%%%%%%%%%%%%%
\section{Projection effects}
\label{app:proj}

We now derive the projection effect contribution to the observed local
small-scale correlation function $\xi_\d(\vr,\tau | h)$.  Here, $\vr$
and $\tau$ are the observationally inferred comoving separation and 
conformal time, respectively.  We make no particular assumptions about
the nature of the tracers which are used to measure the small-scale
correlation function.  We will
only consider the tensor (or vector) case here, so that $h_{00} = 0 = h_{0i}$.  

We begin with Eq.~(45) in \cite{conformalfermi}, which gives
\ba
 \xi(\vr, \tau|h) =\:& \left[1 - a_{ij}(x) r^i \partial_{r}^j + \frac1{\cH} \T(x)\partial_\tau  + 2 c(x)\right] \xi_F(\vr;\tau)
\label{eq:xiproj1}
\ea
where $x$ is the inferred spacetime position of the center of the region in which
the correlation function is measured, and $\xi_F(\vr,\tau)$ is the 
correlation function in the local \fncb~frame.  
Further, $a_{ij}$ is the distortion of the \emph{standard ruler} defined by the correlation function, $\T$ is the shift, in terms of the logarithm
of the scale factor, between constant-proper-time and constant-observed-redshift
surfaces, and $c$ is the perturbation to the observed number density of the
tracer induced by the tensor mode.  We now consider each of these ingredients in turn.  Note that each term in \refeq{xiproj1} is gauge-invariant and in 
principle independently observable.

First, the ruler distortion is most naturally decomposed as
\ba
a_{ij} =\:& \C \nhat_i \nhat_j + \nhat_{(i} \P_{j)k} \B^k + \P_{ik} \P_{jl} \A^{kl},\label{eq:rulerpert}
\ea
where $\P^{ij} = \d^{ij} - \nhat^i\nhat^j$ is the projection operator perpendicular
to the line of sight $\nhat^i$.  
$\C,\,\B_i,$ and $\A_{ij}$ are the gauge-invariant
ruler perturbations defined in \cite{stdruler}, which, when specialized to
a metric with $h_{00} = 0 = h_{0i}$ are given by
\ba
\C =\:&  - \D\ln a 
- \frac12 h_\parallel - \partial_{\chit} \D x_\parallel
\vs
%%%
\B_i =\:& -\P_i^{\  j} h_{jk} \nhat^k  - \nhat^k \partial_{\perp\,i} \D x_k - \partial_{\chit} \D x_{\perp i}
\vs
%%%
\A_{ij} =\:& 
- \D\ln a\: \P_{ij}
- \frac12 \P_i^{\  k}\P_j^{\  l} h_{kl} 
- \frac12 \left(\P_{jk} \partial_{\perp\,i} + \P_{ik} \partial_{\perp\,j}\right) \D x^k\,. \label{eq:coeff}
\ea
The displacements $\D x^i,\,\D\ln a$ are also given in \cite{stdruler} and
again specializing to purely spatial metric perturbations
\ba
\D x_\parallel =\:& - \frac12\int_0^{\chit} d\chi\: h_\parallel
- \frac{1+\zt}{H(\zt)} \D \ln a \label{eq:Dxpar}\\
%%%
\D x_\perp^i =\:& \frac12 \P^{ij} (h_{jk})_o\, \nhat^k \: \chit 
- \int_0^{\chit} d\chi \bigg[
\frac{\chit}{\chi}  \P^{ij} h_{jk}\nhat^k
 -\frac12 (\chit-\chi)\partial_\perp^i h_\parallel\
\bigg]\,.
\label{eq:Dxperp}
\ea
The perturbation to the scale factor at emission is given by
\ba
\D\ln a =\:&  \frac12 \int_0^{\chit} d\chi\: h_{\parallel}' \,.
\label{eq:Dlna}
\ea
In \refeqs{coeff}{Dlna}, metric perturbations outside integrals are evaluated at the source,
unless they are marked by a subscript $o$, in which case they are evaluated at the observer.  Metric perturbations inside integrals are evaluated on the past
lightcone in the background, i.e. at
\be
x^i = \nhat^i \chi ;\quad
x^0 = \tau_o - \chi\,,
\ee
where $\tau_o$ is the conformal time at observation.  Primes denote derivatives with respect to $\tau$, and $\chit \equiv \chib(\zt)$ where $\chib(z)$ is the comoving distance-redshift relation in the background and $\zt$ is the observed redshift.  Further,
\be
\partial_\perp^i = \P^{ij} \partial_j ; \quad
h_{\parallel} = h_{ij} \nhat^i \nhat^j\,.
\ee
The decomposition given by \refeq{rulerpert} allows
us to easily derive the distortions along and perpendicular to the line of 
sight in 3D space.  However, writing the expressions in Cartesian form
leads to more compact expressions.  Using that, for an arbitrary symmetric 
tensor $a_{ij}$,
\be
a_{ij} = a_\parallel \nhat_i \nhat_j + 2 \nhat_{(i} \P_{j)}^{\  k} \nhat^l a_{kl}
+ \P_i^{\  k} \P_j^{\  l} a_{kl}\,,
\ee
\refeq{rulerpert} becomes
\ba
a_{ij} =\:& -\D\ln a\: \d_{ij} - \frac12 h_{ij} - \partial_{(i} \D x_{j)}
= -\frac12 \int_0^{\chit} d\chi\: h_{\parallel}' \: \d_{ij} - \frac12 h_{ij} - \partial_{(i} \D x_{j)}\,.
\label{eq:aij3D}
\ea
The first equality can also be read off directly from Eq.~(30) in \cite{stdruler},
when setting $v^i=0$. Note however, that as discussed in 
\cite{gaugePk,stdruler}, the derivative along the line of
sight is really a derivative with respect to observed redshift, i.e. along the past light cone:
\be
\nhat^i \partial_i \D x^k \equiv \frac{\partial}{\partial\chit} \D x^k = \left(\frac{d\chib(\zt)}{d\zt}\right)^{-1} \frac{\partial}{\partial\zt} \D x^k\,.
\ee
This subtlety is somewhat glossed over in the notation \refeq{aij3D}, which does not make explicit the fundamental difference between line-of-sight and transverse directions.

We further need $\T$, which was derived in \cite{Tpaper} and is given in our case by
\ba
\T =\:& \D\ln a = \frac12\int_0^{\chit} d\chi\: h_{\parallel}'\,.
\ea

Finally, the observed fractional number density perturbation of tracers 
induced by tensor modes was derived in
\cite{GWpaper} (note that $\d z$ in that paper is equal to $\D\ln a$ defined
above).  It is given by
\ba
c =\:& b_e \D\ln a +  \partial_i \D x^i 
+ \Q \:\M_T \,,
\label{eq:ct}
\ea
where  the magnification induced by a tensor mode is
\be
\M_T =
- 2\D\ln a +\frac{1}{2}h_\parallel
- \frac{2\Delta x_\parallel}{\chit} + 2 \hat\kappa\,,
\label{eq:MT}
\ee
and the convergence is 
\ba
\hat\k \equiv\:& -\frac{1}{2}\partial_{\perp i}\Delta x_\perp^i 
= \frac54 h_{\parallel o} -\frac12 h_{\parallel}
 - \frac12 \int_0^{\chit} d\chi \Big[ 
h_\parallel' + \frac3{\chi} h_\parallel \Big] 
- \frac14 \nabla^2_\Omega \int_0^{\chit} d\chi \frac{\chit-\chi}{\chit\:\chi} 
h_\parallel. \label{eq:kappa}
\ea
Here $\nabla^2_\Omega = \chit^2 \nabla_\perp^2$ denotes the Laplacian on the unit
2-sphere.  The Jacobian in \refeq{ct} is then given by
\be
\partial_i \D x^i  = \partial_{\chit} \Delta x_\parallel
+ \frac{2\Delta x_\parallel}{\chit} - 2 \hat\kappa \,.
\label{eq:jacobian}
\ee
Note again the subtlety in the Cartesian notation.  The number density
modulation is governed by two tracer-dependent parameters: the 
magnification bias parameter $\Q$, given in the simplest case of a sharp
flux-limited survey by $\Q = - d\ln \bar{n}_g/d\ln f_{\rm cut}$;  and the
paramater $b_e$, which quantifies the redshift evolution of the comoving
number density of tracers through
\be
b_e \equiv \frac{d\ln (a^3\bar n_g)}{d\ln a}\Big|_{\!\zt} 
= - (1+\zt)\frac{d\ln (a^3\bar n_g)}{dz}\Big|_{\!\zt} \,.
\label{eq:btdef}
\ee

Putting everything together, the observed local two-point correlation
function becomes
\ba
 \xi(\vr, \tau|h) = \Bigg[ & 1 
+ \left\{\frac12 \int_0^{\chit} d\chi\: h_{\parallel}' \: \d_{ij} + \frac12 h_{ij} + \partial_{(i} \D x_{j)}\right\}  r^i \partial_{r}^j 
+ \frac1{2\cH} \int_0^{\chit} d\chi\: h_{\parallel}' \:\partial_\tau  \vs
%%%
& + 2 \left\{ \frac{b_e}2\int_0^{\chit} d\chi\: h_{\parallel}' + \partial_i \D x^i 
+ \Q \:\M_T \right\}\Bigg] \xi_F(\vr;\tau)\,.
\label{eq:xiproj2}
\ea

\end{widetext}

%%%%%%%%%%%%%%%%%%%%%%%%%%%%%%%%%%%%%%%%%%%%%%%%%%%%%%%%%%%%%%%%%%%%%%%%%%%
%%%%%%%%%%%%%%%%%%%%%%%%%%%%%%%%%%%%%%%%%%%%%%%%%%%%%%%%%%%%%%%%%%%%%%%%%%%
%\bibliographystyle{arxiv_physrev}
\bibliography{GW}

\end{document}